\documentclass{WileyMSP-template}

\usepackage{amsmath}
\usepackage{amsthm}
\usepackage{amssymb}
\usepackage{xcolor}
\usepackage{hyperref}
\usepackage{textcomp}
\usepackage{cite}
\usepackage[normalem]{ulem}

\begin{document}

\rhead{\includegraphics[width=2.5cm]{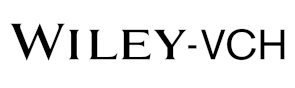}}

\title{Exact non-Gaussian {\sout{stationary}} statistics of a trapped Brownian particle
subject to Poisson-shot noise in an active bath}

\title{Brownian particle in a Poisson-shot-noise  active bath:\\  exact statistics, effective temperature, and  inference}

\maketitle

\author{Costantino Di Bello$^\dagger$\,$^1$}, 
\author{Rita Majumdar$^\dagger$\,$^{2,3}$\let\thefootnote\relax \footnote{$\dagger$ : equal contribution}}, 
\author{Rahul Marathe\,$^2$}, 
\author{Ralf Metzler\,$^{1,4}$}
and\author{\'Edgar Rold\'an\,$^3$},

\begin{affiliations}
$^1$ Institute of Physics \& Astronomy, University of Potsdam, 14476 Potsdam, Germany\\

$^2$ Department of Physics, Indian Institute of Technology, Delhi, Hauz Khas 110016,
New Delhi, India\\
$^3$ ICTP -- The Abdus Salam International Centre for Theoretical Physics,  34151
Trieste, Italy\\
$^4$ Asia Pacific Centre for Theoretical Physics, Pohang 37673, Republic of Korea

\end{affiliations}


\begin{abstract}
We study the dynamics of an overdamped Brownian particle in a thermal bath that contains a dilute solution of active particles. The
particle moves in a harmonic potential and experiences  Poisson shot-noise kicks with specified amplitude
distribution due to moving active particles in the bath. From the Fokker-Planck
equation for the particle dynamics we derive the stationary solution for the
displacement distribution along with the moments characterising mean, variance,
skewness, and kurtosis, as well as finite-time first and second moments. We also compute  an effective temperature through the fluctuation-dissipation theorem and show that equipartition theorem holds for all zero-mean kick distributions, including those leading to non-Gaussian stationary statistics.
For the case of Gaussian-distributed active kicks we find
a re-entrant behaviour from non-Gaussian to Gaussian  stationary states and a heavy-tailed
leptokurtic distribution across a wide range of parameters  as seen in recent
experimental studies. Further analysis reveals statistical signatures of the irreversible dynamics of the particle displacement in terms of the time asymmetry of cross-correlation functions. Fruits of our work is the development of a compact inference scheme that may allow experimentalists to extract the rate and moments of underlying shot-noise solely from the statistics the particle position.
\end{abstract}

\section{Introduction}

Following the probabilistic description of Brownian motion by Einstein and
Smoluchowski \cite{einstein,smolu}, Langevin introduced the concept of the
fluctuating force \cite{langevin}, to capture (schematically \cite{levy})
the motion of a single particle. Fluctuating forces have meanwhile become
a key principle in the formulation of non-equilibrium statistical physics
\cite{zwanzig,brenig,toda}. We note that the classically considered systems
are connected to a thermal bath, effecting, inter alia, the temperature
dependence of the diffusion coefficient of a Brownian particle embodied in
the Einstein-Smoluchowski relation \cite{vankampen}.
\vspace{0.5cm}\\
The search for fundamental laws governing stochastic processes in a
non-equilibrium system is one of the most active fields of research within
statistical physics. While thermodynamic laws are well established in the
context of equilibrium macroscopic systems \cite{landau}, their understanding
turns out to be significantly more challenging when studying the erratic motion
of microscopic non-equilibrium systems, such as unicellular organisms, which
lead to fluctuating transfer of energy and matter.  Stochastic thermodynamics
is currently playing a central role in establishing a theoretical framework
to study small systems far from equilibrium, in which fluctuations and
randomness play a significant role \cite{sekimoto98,Seifert12}.\vspace{0.5cm}\\

A key goal of contemporary non-equilibrium thermodynamics is to find universal
principles governing the behaviour of active matter. Active matter has recently
attracted considerable attention in statistical physics, biophysics, and soft
matter \cite{Vicsek12,Ramaswamy10,Volpe16}. Popular toy-models in statistical
physics of active matter are the so-called (microscopic) active particles, which
exhibit self-propulsion in fluctuating media by consuming and dissipating
internal and environmental sources of energy. Such models have been very
successful in describing experimental records of the motion of, e.g.,
bacterial suspensions of different types of bacteria or light or chemical
gradient controlled artificial micro-swimmers \cite{Volpe16,Krishnamurty16}.
Active systems operate away from equilibrium and thus do not satisfy classical
detailed balance nor a fluctuation-dissipation relation---but follow recently
discovered principles derived in the framework of stochastic thermodynamics
\cite{leticia,Maes14,Maes15,Cates,Genesotto18,Edgar}. It is customary to describe the motion
of active particles using overdamped Langevin equations, in which an active
noise component is considered along with the  thermal Gaussian white noise,
giving rise to non-trivial statistics \cite{Ramaswamy10,Volpe16}. Examples
include Ornstein-Uhlenbeck noise, run-and-tumble motion, telegraphic, L\'evy,
and Poisson shot noise \cite{spiechowicz13,Basu20,Marathe18,Marathe19,Marathe22,
Ruben23}. In general, it is not possible to derive exact analytical expressions
for the emerging statistics (e.g., finite-time moments, stationary distribution,
etc.) of Langevin equations with active noise, with the exception of few examples
for which such calculations becomes a formidable task \cite{Abhishek21,Lucente23,
Shee22}.\vspace{0.5cm}\\

A fundamental question in statistical physics is the interaction of thermal
energy and confinement. At equilibrium, the particle displacement in a
conservative force field is described by the Boltzmann distribution
\cite{landau,vankampen}. Of particular interest in physics is the linear
Hookean force field. In a thermal bath this corresponds to the Brownian
harmonic oscillator \cite{brenig}. For colloidal particles a linear force
can be implemented experimentally by optical tweezers \cite{lenerev}. While
the relaxation dynamics of the colloidal particle towards equilibrium may be
non-exponential in complex fluids \cite{lene1}, the equilibrium distribution
typically remains Gaussian in a thermal bath, with an externally tunable
width depending on the force constant of the tweezers trap.\vspace{0.5cm}\\

Optical tweezers setups can also be immersed in active reservoirs. Such studies
showed that the displacement distribution of the confined particle becomes
progressively non-Gaussian with increasing activity of the (bacterial) bath
\cite{Blickle12,Martinez16,Krishnamurty16,Wu20,Krishnamurty21,Albay21,Cheng22}.
The distribution of the system exhibits a concentrated central region with
heavier-than-Gaussian tails, thereby enhancing the system dynamics
\cite{Krishnamurty21}.\footnote{We note that non-Gaussian tails have also been
shown for the unconfined active motion of polymers in active particle baths
\cite{jaeoh}, microswimmers \cite{activebath2}, social amoeba \cite{carsten},
self-propelling Janus particles \cite{lowen}, progenitor cells \cite{runtumble},
or nematodes \cite{hapca}, for which distributed-parameter models have been
discussed \cite{klapp,elisabeth}.} Under the influence of coloured noise, two
approximations to the stationary probability distribution for Langevin dynamics
with coloured  Ornstein-Uhlenbeck noise in conservative potentials are
extensively investigated in literature (Fox \cite{Fox86} and
UCNA~\cite{Jung87}). Recently, several approximations have been introduced
for stochastic systems that relax to a non-equilibrium stationary state
in the presence of coloured noise. In a one-dimensional active system,
the precise stationary probability distribution is evaluated rarely,
e.g., for run-and-tumble dynamics \cite{Cates09}. In a system composed of
interacting active particles, the explicit formula for the non-equilibrium
stationary probability distribution is achieved using the unified
coloured noise approximation (UCNA) \cite{Maggi15} as well as Fox's
approximation \cite{Aswin21}. For active-Ornstein-Uhlenbeck particles (AOUPs),
the steady-state distribution is calculated at a small but finite persistence
time. The particle shows a non-Boltzmann distribution but still maintains
detailed balance \cite{Fodor16}.\vspace{0.5cm}\\

A {plausible model for the noise exerted by a dilute solution of active systems (e.g. bacteria) to an optically-trapped colloidal particle is a sequence of    active kicks with  arrivals  at random Poissonian
times. Such noise has also been used to describe recent experiments  in soft~\cite{Krishnamurty21,Pak20} and granular matter~\cite{Lucente23}. } Analytically solving
the Fokker-Planck equation with such a form of non-Gaussian active noise
poses a considerable challenge. Thus, in this context, we here address
the Fokker-Planck equation equation for a linear stochastic model in the
presence of both thermal (Gaussian white) and active Poisson shot noise
(PSN), aiming to calculate the stationary probability density, along with
closed-form solutions for its moments. PSN has recently captured the
interest of the statistical physics community due to its rich and complex phenomenology \cite{vandenbroeck,spiechowicz13,bialas23,kanazawaPRL, kanazawaJSP,Lucente23}. One of the key results of this research line was the derivation of the Fokker-Planck equation associated with the Langevin description of stochastic models
subject to PSN. In \cite{vandenbroeck} it was shown how PSN can be seen as
the limit of a dichotomous Markov process. In particular, it was observed that PSN can significantly affect the statistical properties of a particle in a thermal bath subjected to a periodic potential, either inducing absolute
negative mobility \cite{spiechowicz13}, or enhancing its transport properties \cite{spiechowicz14,spiechowicz15,bialas23}. Reference \cite{kanazawaPRL} discusses the non-Gaussianity of this process and shows how the microscopic rate of jumps is connected to the macroscopic stationary probability of the process. The same authors discuss in reference \cite{kanazawaJSP} how to solve the model with an arbitrary non-linear frictional force.
It was also shown \cite{bialas20} how a system with PSN can exhibit a Brownian yet non-Gaussian behavior having a
mean squared displacement increasing with time while possessing a non-Gaussian distribution of the position (see also \cite{prx} for more details). It has
also been found \cite{baule} that a particle in a bath with PSN can have a much more efficient escape rate with respect to the case of pure Gaussian white noise. These results were illustrated with efficient methods of simulating stochastic differential equations with PSN in \cite{numerical07,numerical09}.\vspace{0.5cm}\\

In this paper we consider a minimal Langevin dynamics model for an overdamped Brownian
particle that is confined in a harmonic potential and immersed in diluted solution of active systems, i.e. the particle is simultaneously subject to
 a thermal and an active noise, the latter being modelled by a PSN. The PSN here implies that in an infinitesimal time interval $dt$ there is a probability $rdt$ that the particle receives a shot, or kick, instantaneously
shifting its position from $x(t)$ to $x(t+dt)=x(t)+\text{shot}$. Each kick
is considered to be independent of the position at time $t$, and all kicks
are independent identically distributed (i.i.d.) random variables. The
intensity of the kicks may come from some specified probability density
function (PDF). For simplicity, we have assumed instantaneous kicks, see e.g. Ref.~\cite{hanggi2} for a generalization to PSN with finite duration pulses. 
The mathematical formulation of such a stochastic process was discussed in \cite{hanggi78,hanggi80}, and its steady-state distribution was obtained in \cite{eliazar_klafter}.\vspace{0.5cm}\\

The rest of this work is organised as follows. In Sec.~\ref{sec:model}, we introduce our setup, a Langevin equation describing the motion of a trapped Brownian particle that is subject to Gaussian white and Poisson shot noises, and we establish the Fokker-Planck equation (FPE) associated with its dynamics. In Sec.~\ref{sec:main_results} we derive exact analytical expressions for the stationary PDF and the moments of the particle position from the FPE. To this aim we solve the FPE in the rather general setting in which the amplitude of the PSN is generated from a specified distribution. Next, we link our model to a possible experimental scenario, providing an exact inference method for the PSN kick statistics. Then we derive general results related to the effective temperature for our model in terms of the Onsanger regression principle. In Sec.~\ref{sec:Gaussian_kicks} we consider a specific case of the Gaussian distribution of the PSN and derive analytical results. For this example, we discuss the notion of the effective temperature and provide insights about the non-Gaussianity of the process through a detailed analysis of the excess kurtosis. We conclude in Sec.~\ref{sec:conclusion} with some remarks in relation with recent soft-matter experiments with active matter, and provide insights about irreversibility and dissipation of our model. Details of the derivations, like calculations of two non-stationary moments and long-time susceptibility, and details of numerical simulations are relegated to the appendices.

\section{Model}
\label{sec:model}

We consider a one-dimensional overdamped Brownian particle trapped in the
harmonic potential $U(x)=\kappa x^2/2$ of force constant $\kappa$ that is
in simultaneous contact with a thermal bath and an active bath. The
overdamped Langevin equation describing the motion of the Brownian particle
then reads
\begin{equation}
\label{eqn:Langevin}
\gamma\dot{X}_t=-\kappa X_t+\sqrt{2k_BT\gamma}\xi_t+\gamma\eta_t,
\end{equation}
where $X_t$ is the particle position at time $t$, $\gamma$ is the friction
coefficient, $k_B$ is the Boltzmann constant, and $T$ is the temperature
of the thermal bath. The stochastic force $\xi_t$ is a Gaussian white noise
with zero mean $\langle\xi_t\rangle=0$ and autocorrelation $\langle\xi_t\xi
_{t'}\rangle=\delta(t-t')$. The term $\eta_t$ is a stochastic force, which
models instantaneous random displacements experienced by the particle as a
result of a kick received from constituent particles of an active bath. We
assume that these kicks occur with Poissonian waiting times with constant
rate $\omega$, i.e., $\eta_t$ is a PSN that can be written as \cite{hanggi80}
\begin{equation}
\eta_t=\sum_{i=1}^{N_t}Y_i\delta(t-t_i).
\end{equation}
Here, $Y_i$ is the positional displacement experienced by the particle due
to the $i$th active kick. Furthermore, $t_i$ are the arrival times of a
Poisson counting process with rate $\omega$, and $N_t$ is the total number
of kicks occurring up to time $t$. Hence, the PDF of $N_t$, $P(n,t)=P(N_t=n)$,
is Poissonian and given by
\begin{equation}
P(n,t)=\exp(-\omega t)\dfrac{(\omega t)^{n}}{n!}.
\end{equation}
As mentioned, we assume that the sequence $Y_i$ of kick amplitudes is an i.i.d.
process with dimension of length in which each $Y_i$ is sampled from a predefined
PDF $\rho_a(y)$. Under these assumptions, it follows that the process $\eta_t$
has the following statistical features 
\begin{equation}
\label{eqn:noise_eta}
\langle\eta_t\rangle=\omega\langle Y\rangle_a,\quad
\langle\eta_t\eta_s\rangle-\langle\eta_t\rangle\langle\eta_s\rangle=\omega
\langle Y^2\rangle_a\delta(t-s),
\end{equation}
where $\langle Y^n\rangle_a$ denotes the $n$th moment of the PDF $\rho_a$.
Moreover, we also assume that the two noises $\xi_t$ and $\eta_t$ are
independent, i.e., $\langle\xi_t\eta_s\rangle=\langle\xi_t\rangle\langle
\eta_s\rangle=0$. We finally introduce the characteristic relaxation time
\begin{equation}
\tau\equiv\frac{\gamma}{\kappa},
\end{equation}
which allows us to rewrite Eq.~\eqref{eqn:Langevin} in the more convenient
form
\begin{equation}
\tau\dot{X}_t=-X_t+\sqrt{2\tau\frac{k_BT}{\kappa}}\xi_t+\tau\eta_t.
\label{eqn:Langevin2}
\end{equation}

We may decompose the Langevin equation \eqref{eqn:Langevin2} into the two separate component stochastic processes $X_{1,t}$ and $X_{2,t}$ such that
$X_t=X_{1,t}+X_{2,t}$, allowing us to write
\begin{eqnarray}
\tau\dot{X}_{1,t}&=&-{X}_{1,t}+\sqrt{2\tau\frac{k_BT}{\kappa}}\xi_t,\\
\tau\dot{X}_{2,t}&=&-X_{2,t}+\tau\eta_t.
\end{eqnarray}
The initial conditions are such that $X_0=X_{1,0}+X_{2,0}$. The formal
solutions of the Langevin equations for $X_{1,t}$ and $X_{2,t}$ read
\begin{equation}
\begin{aligned}
X_{1,t}&=\exp(-t/\tau)\left[X_{1,0}+\sqrt{\frac{2k_BT}{\tau\kappa}}\int_0^tds
\xi_s\exp(s/\tau)\right],\\
X_{2,t}&=\exp(-t/\tau)\left[X_{2,0}+\int_0^tds\eta_s\exp(s/\tau)\right].
\end{aligned}
\end{equation}
Thus, after summation we get
\begin{equation}
\label{eqn:Langevin_solution}
X_t= \exp(-t/\tau) \left[X_0+\int_0^tds \left(\sqrt{\frac{2k_BT}{\tau\kappa}}\xi_s +\eta_s\right) \exp(s/\tau)\right],
\end{equation}
which is precisely the solution of \eqref{eqn:Langevin}. Clearly, this
factorisation also holds in the stationary state. By denoting with $X=
\lim_{t\to\infty}X_t$, with $X_1=\lim_{t\to \infty}X_{1,t}$ and $X_2
=\lim_{t\to\infty}X_{2,t}$ we obtain that $X=X_1+X_2$. Since $X$ is the
sum of two independent random variables, we can use some simple identities
to compute its moments, variance, skewness and kurtosis, as we will show
in Sec.~\ref{sec:main_results}.

\begin{figure}
\begin{center}
\includegraphics[width=7.0cm,height=4.0cm,angle=0]{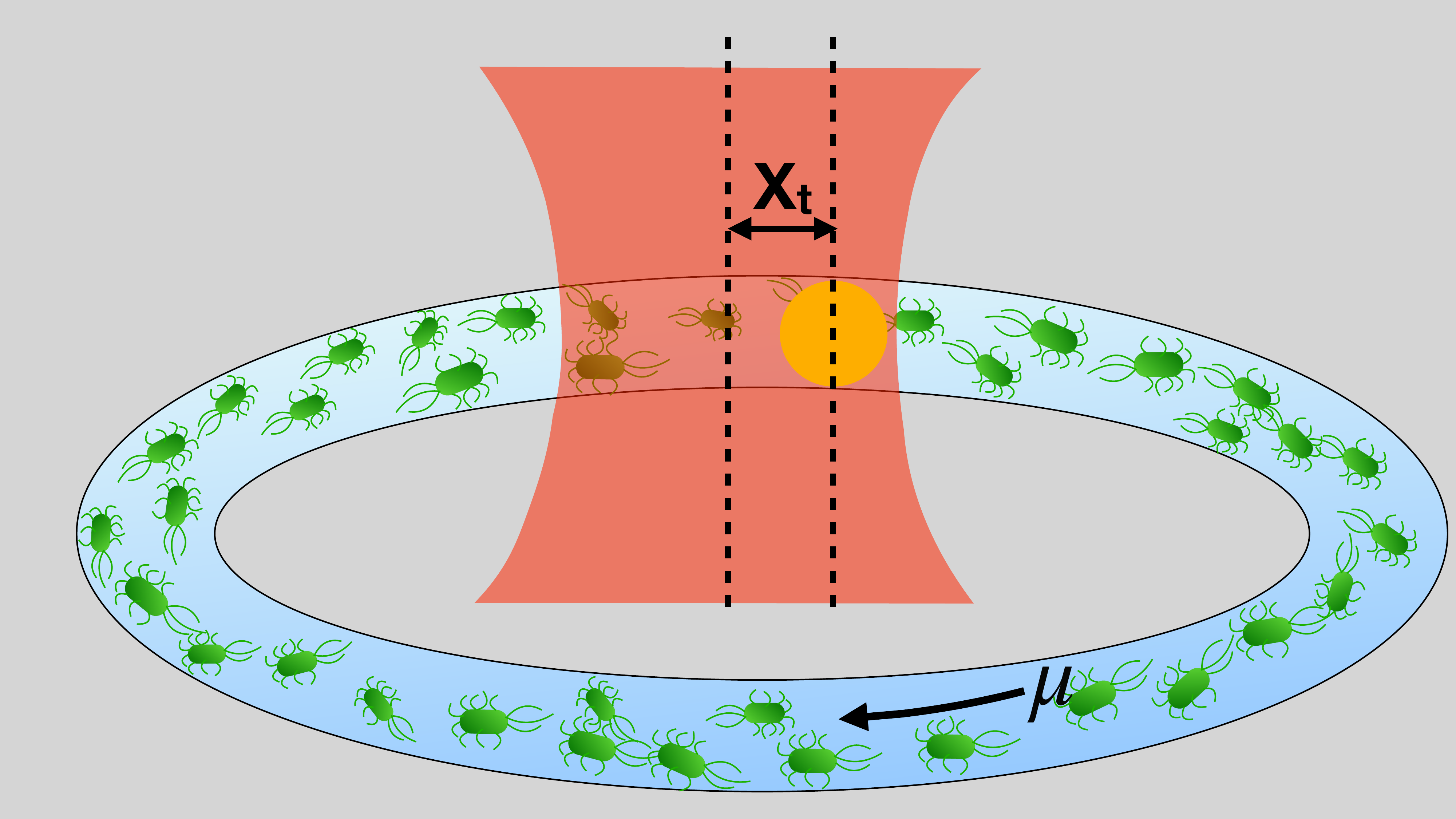} \hspace{0.15cm}
\includegraphics[width=4.5cm,height=4.0cm,angle=0]{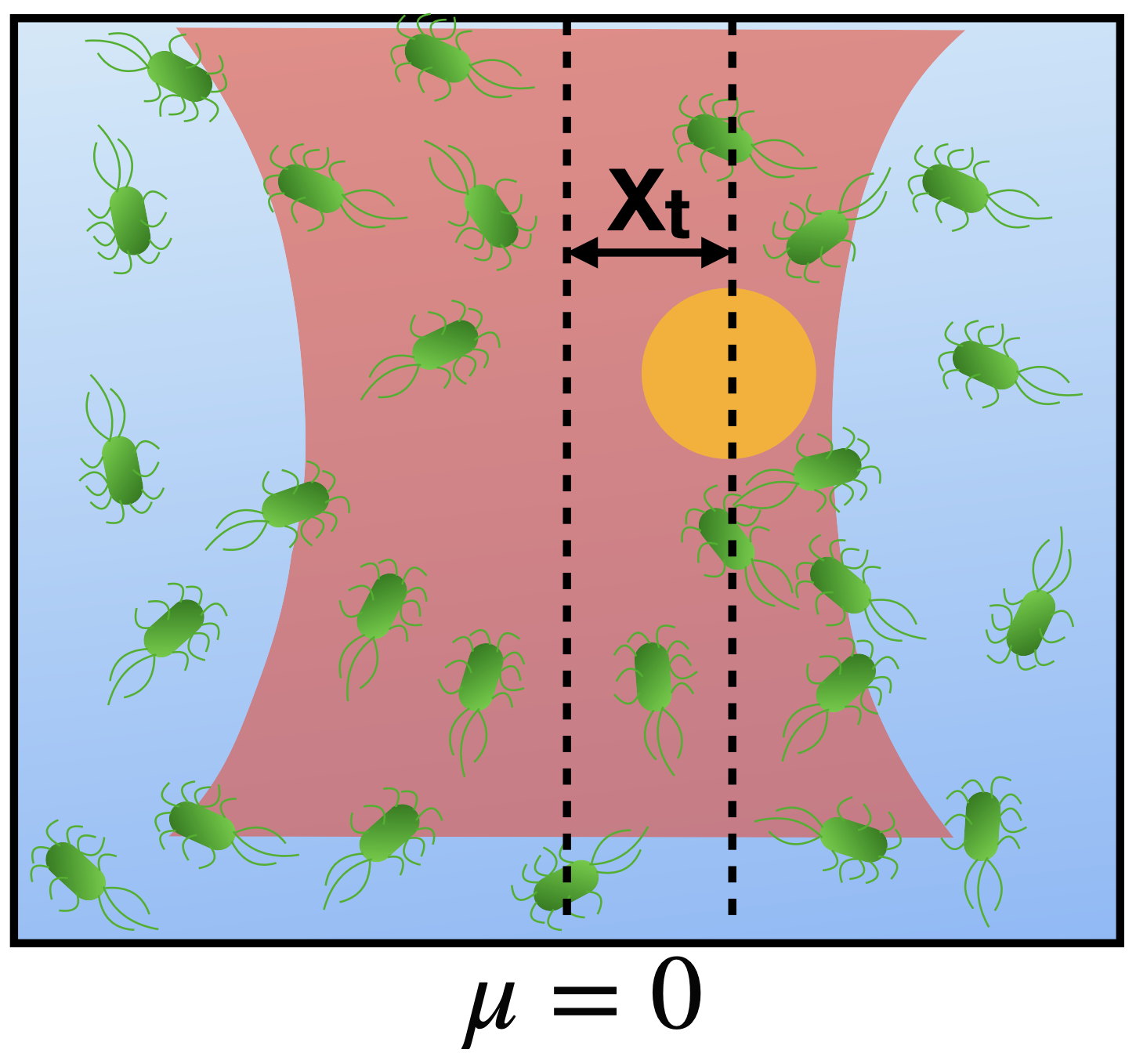} \hspace{0.15cm}
\includegraphics[width=6.0cm,height=4.0cm,angle=0]{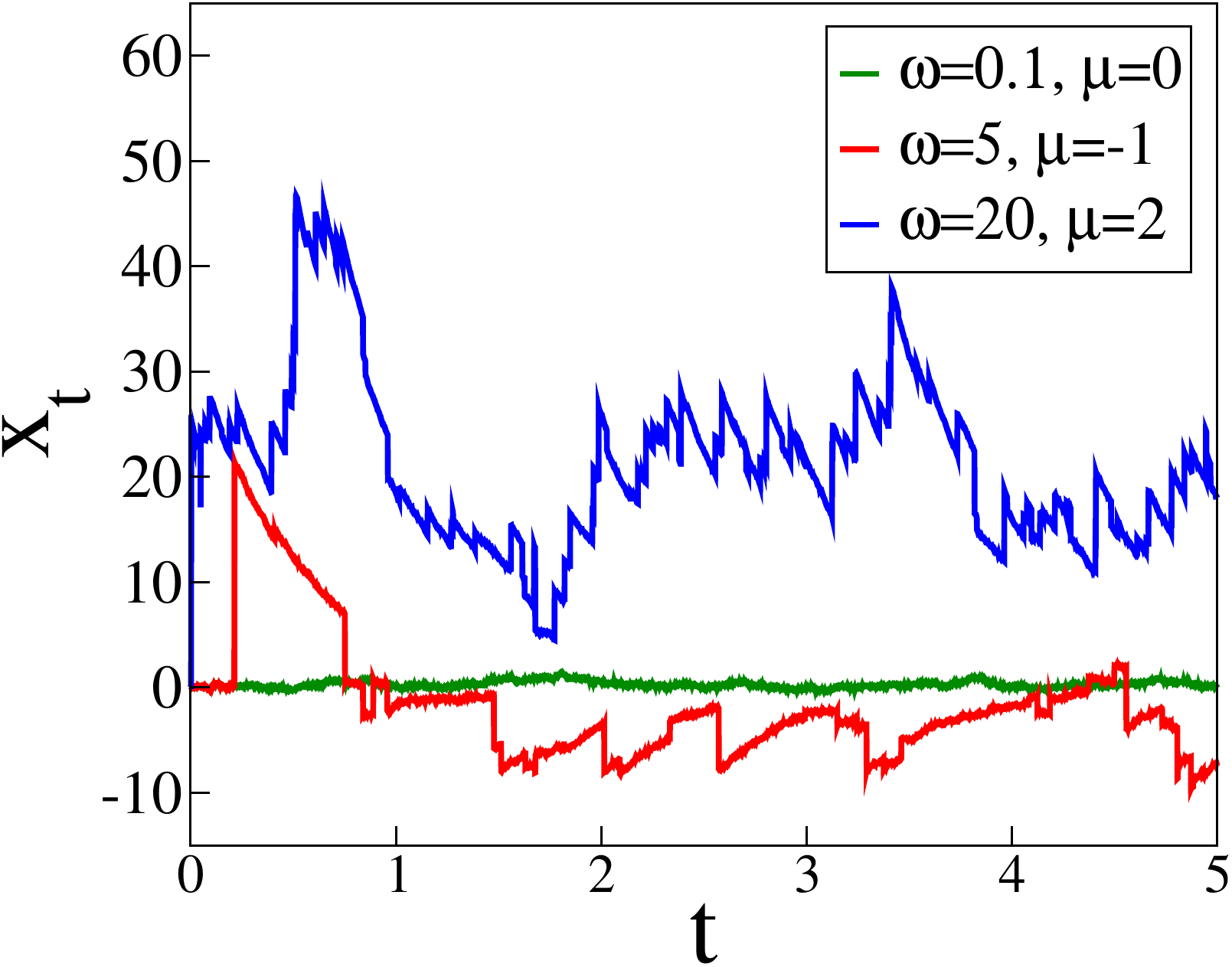} 
\end{center}
\caption{(Left panel) Sketch of a Brownian particle trapped  with
an optical tweezer within periodic boundary conditions in a dilute solution of active particles with a nonzero net average drift ($\mu\neq 0$). (Centre Panel) The trapped Brownian particle is subject to closed boundary conditions and 
immersed in a dilute active bath composed of active particles with zero
net drift ($\mu=0$). (Right Panel) Representative stochastic trajectories
of the Brownian particle in the active bath exerting a Poisson shot noise (PSN) on the particle (on top of the thermal noise). In this illustration, the amplitude of the kicks of the PSN is drawn from a  Gaussian distribution given by Eq.~\ref{eqn:rho_Gaussian}. Different colors represent trajectories for different parameter values varying the rate  rate $\omega$ and mean value of the amplitude $\mu$ of the PSN kicks (see legend). }
\label{schematic_diagram}
\end{figure}

Figure~\ref{schematic_diagram} sketches a possible experimental realisation
of our model in which a colloidal particle is embedded in a Newtonian fluid
at temperature $T$ and trapped in a static harmonic potential created, e.g.,
by optical tweezers. The colloid is put in contact with a non-equilibrium bath
of self-propelled bacteria. The bacterial activity may be tuned externally
(e.g., by setting a constant concentration of nutrients) in such a way that both
the kick rate $\omega$ and the amplitude distribution $\rho_a(Y)$ of the kicks
exerted on the particle can be maintained at a steady level. Furthermore, the
particle and bacteria may be trapped in a periodic chamber where fluid flows
at a constant speed giving rise to an asymmetric bacterial kick distribution
$\langle{Y}\rangle_a\neq0$ (Fig.~\ref{schematic_diagram}, left panel), or in
a closed chamber for which the bacterial kick distribution is symmetric around
zero $\langle{Y}\rangle_a=0$ (Fig.~\ref{schematic_diagram}, middle panel).

The colloid dynamics shows some exciting behaviour as we change the mean spiking
rate $\omega$ or the bacterial activity. Here, we present some trajectories
of the particle driven by the active noise (Eq.~\ref{eqn:rho_Gaussian}
and Sec.~\ref{sec:Gaussian_kicks}) for different $\omega$. It is evident
that for small $\omega$ the average value of the position of the colloid
is close to zero, and it behaves like a Gaussian process. As we increase
the rate $\omega$, the jumps in the trajectory become prominent, and the
system is far from equilibrium. We also varied the mean velocity of the
bacteria. The green trajectory corresponds to $\mu=0$, when a particle is
optically trapped in a closed container filled with bacteria.  Similarly,
the red trajectory stands for the average velocity $\mu=-1$, and the blue
trajectory is for $\mu=2$, representing an optically-trapped particle in a
bath with bacteria that experience a homogeneous fluid flow.

It was shown previously \cite{hanggi78} that the above system can be described
by the Fokker-Planck equation
\begin{equation}
\label{eqn:FokkerPlanck}
\tau\dfrac{\partial}{\partial t}P(x,t)=\dfrac{\partial}{\partial x}\left[x
P(x,t)\right]+\frac{k_BT}{\kappa}\dfrac{\partial^2}{\partial x^2}P(x,t)+
\omega\tau\int_{-\infty}^{\infty}[P(x-y,t)-P(x,t)]\rho_a(y)dy,
\end{equation}
where $P(x,t)=\mathrm{Prob}\left\{ x\leq X_t<x+dx\right\}$ denotes the
probability density over space at time $t$, with some initial condition
$P(x,0)=P_0(x)$ and with vanishing probability at the extremities, $\lim_{
x\to\pm\infty}P(x,t)=0$. We will not focus on the time-dependent probability
$P(x,t)$ but rather its stationary distribution $P(x)\equiv\lim_{t\to\infty}
P(x,t)$. We will use Eq.~\eqref{eqn:FokkerPlanck} to derive $P(x)$. We will
proceed by deriving the analytical expression for $P(x)$ with its first and
second moments, skewness, and kurtosis, for arbitrary kick intensity distribution $\rho_a(y)$. We will then consider the specific case
of Gaussian kick intensities, and conclude with some results concerning other
intensity distributions and with some future directions of research.

\section{Main results}
\label{sec:main_results}

We now derive the stationary PDF $P(x)$. To this end, rewrite the FPE Eq.
\eqref{eqn:FokkerPlanck} in the form
\begin{equation}
0=\dfrac{\partial}{\partial x}\left[xP(x)\right]+\frac{k_BT}{\kappa}\dfrac{
\partial^2}{\partial x^2}P(x)-\omega\tau P(x)+\omega\tau\int_{-\infty}^{\infty}
P(x-y)\rho_a(y)dy,
\end{equation}
where we set the left hand side to zero and used the normalization condition
of $\rho_a$. We notice that the integral term in this integro-differential
equation is just a convolution of two functions, and it is then convenient to
apply a Fourier transform. We denote the respective Fourier transform of $P(x)$
and $\rho_a(y)$ by $\hat{P}(q)=\int_{-\infty}^{\infty}\exp(\mathrm{i}qx)P(x)dx$
and $\hat{\rho}_a(q)=\int_{-\infty}^{\infty}\exp(\mathrm{i}qy)\rho_a(y)dy$,
where $\mathrm{i}$ is the imaginary unit. We then obtain
\begin{equation}
0=-q\dfrac{\partial}{\partial q}(\hat{P}(q))-\frac{k_BT}{\kappa}q^2\hat{P}(q)+
\omega\tau\hat{P}(q)\hat{\rho}_a(q)-\omega\tau\hat{P}(q),
\end{equation}
which is an ordinary differential equation with the boundary condition $\hat{P}
(0)=1$. After some simple algebra we obtain
\begin{equation}
q\dfrac{\partial\hat{P}(q)}{\partial q}=\left(-\frac{k_BT}{\kappa}q^2+\omega
\tau\hat{\rho}_a(q)-\omega\tau\right)\hat{P}(q),
\end{equation}
from which, in turn, we find
\begin{equation}
\dfrac{\partial\ln(\hat{P}(q))}{\partial q}=-\frac{k_BT}{\kappa}q+\omega\tau
\left(\dfrac{\hat\rho_a(q)-1}{q}\right).
\end{equation}
Hence the stationary PDF in Fourier domain reads
\begin{equation}
\label{eqn:stat_P_fourier}
\hat{P}(q)=\exp\left(-\frac{k_BT}{2\kappa}q^2\right)\exp\left(\omega\tau\int_0^q
\dfrac{\hat{\rho}_a(q')-1}{q'}dq'\right).
\end{equation}
As we can see the previous expression is the product of the two terms
\begin{equation}
\label{eqn:p1_p2_def}
\hat{P}_1(q)=\exp\left(-\frac{k_BT}{2\kappa}q^2\right),\quad
\hat{P}_2(q)=\exp\left(\omega\tau\int_0^q\dfrac{\hat{\rho}_a(q')-1}{q'}dq'
\right),
\end{equation}
where $\hat{P}_1(q)$ is the characteristic function of the random variable
$X_1$, while $\hat{P}_2(q)$ is the characteristic function of $X_2$. Therefore,
as stated before, this confirms that $X=X_1+X_2$, since the characteristic
function of the sum of two random variables is the product of the two
characteristic functions. Finally, by inverting the Fourier transform, $P(x)$
reads
\begin{equation}
\label{eqn:stat_P}
P(x)=\int_{-\infty}^{\infty}\dfrac{dq}{2\pi}\exp\left(-\mathrm{i}qx\right)\exp
\left(-\frac{k_{\rm B}T}{2\kappa}q^2+\omega\tau I(q)\right),
\end{equation}
where
\begin{equation}
\label{eqn:I_definition}
I(q)\equiv\int_0^q\dfrac{\hat{\rho}_a(q')-1}{q'}dq'.
\end{equation}
% \sout{Equation \eqref{eqn:stat_P_fourier} (or equivalently \eqref{eqn:stat_P}) is the first main result of the paper.} 
Equation \eqref{eqn:stat_P_fourier} was already derived in Ref.~\cite{eliazar_klafter} through an alternative mathematical approach. For a given choice for $\rho_a$ Eq.~\eqref{eqn:stat_P} can be numerically implemented using the GNU Scientific Library (GSL), see Appendix~\ref{sec:code} for further details. In the next subsection
we will discuss the moments of the random variables $X$, $X_1$, and $X_2$.

\subsection{Moments of the distribution}
\label{sec:moments}

We now derive the analytical expressions for the lower-order moments of $X$ in
the stationary state. We use the notation
\begin{equation}
\langle X^n\rangle=\int_{-\infty}^{\infty}x^nP(x)dx
\label{eqn:X_n}
\end{equation}
for the $n$th moment of the random variable $X$. In what follows, we will
mainly focus on the variance $\mathrm{Var}[X]$, skewness $\Tilde{\mu}_3[X]$,
and excess kurtosis $\mathcal{K}_{\rm ex}[X]$ of the random variable $X$. The
general expressions for these central statistical quantities, in terms of the
central moments of $X$, are given by
\begin{equation}
\mathrm{Var}[X]\equiv\langle(X-\langle X\rangle)^2\rangle,\quad 
\Tilde{\mu}_3[X]\equiv\dfrac{\langle(X-\langle X\rangle)^3\rangle}{\mathrm{Var}
[X]^{3/2}},\quad\mathcal{K}_{\rm ex}[X]\equiv\frac{\langle (X-\langle X \rangle)
^4\rangle}{\mathrm{Var}[X]^{2}}-3.
\label{skew_kurt_formula}
\end{equation}
The statistical quantities defined in Eq.~\eqref{skew_kurt_formula} may be
directly evaluated by expanding the binomials and using the identity 
\begin{equation}
\label{eqn:moments_identity}
\langle X^n\rangle=\dfrac{1}{\mathrm{i}^n}\dfrac{d^n}{dq^n}\hat{P}(q)\biggr
\rvert_{q=0},
\end{equation}
where $\hat{P}(q)$ is given by Eq.~\eqref{eqn:stat_P_fourier}. After some
algebra, we obtain the first four moments of the particle position in the
stationary state,
\begin{eqnarray}
&&\hspace{-1cm}\langle X\rangle=\omega\tau\langle Y\rangle_a,\label{eqn:<X>}\\ 
&&\hspace{-1cm}\langle X^2\rangle=\frac{k_BT}{\kappa}+\frac{\omega\tau}{2}
\langle Y^2\rangle_a+\omega^2\tau^2\langle Y\rangle^2_a,\label{eqn:X_2}\\ 
&&\hspace{-1cm}\langle X^3\rangle=\frac{\omega\tau}{3}\langle Y^3\rangle_a+
\omega^3\tau^3\langle Y\rangle_a^3+3\omega \tau\langle Y\rangle_a\left(
\frac{k_BT}{\kappa}+\frac{\omega\tau}{2}\langle Y^2\rangle_a\right),
\label{eqn:X_3}\\
\nonumber
&&\hspace{-1cm}\langle X^4\rangle=\frac{\omega\tau}{4}\langle Y^4\rangle_a+
\frac{4\omega^2\tau^2}{3}\langle Y\rangle_a\langle Y^3\rangle_a+\omega^4\tau
^4\langle Y\rangle_a^4+6\omega^2\tau^2\langle Y\rangle^2_a\left(\frac{k_BT}{
\kappa}+\frac{\omega\tau}{2}\langle Y^2\rangle_a\right)\\
&&+3\left(\frac{k_BT}{\kappa}+\frac{\omega\tau}{2}\langle Y^2\rangle_a\right)^2.
\label{eqn:X_4}
\end{eqnarray}
Substituting expressions (\ref{eqn:<X>}-\ref{eqn:X_4}) into
Eq.~\eqref{skew_kurt_formula} we directly find the analytical expressions for
the stationary variance, skewness, and excess kurtosis associated with
the particle position.

We mention an alternative yet insightful way to evaluate the moments of $X$
through the statistics of $X$. As mentioned, in the stationary state $X=X_1
+X_2$, with $X_1$ and $X_2$ being independent random variables. This property
implies the compact expressions of the moments
\begin{equation}
\label{eqn:moments_sum}
\langle X^n\rangle=\sum_{m=0}^n\binom{n}{m}\langle X_1^m\rangle\langle X_2^{n-m}
\rangle,
\end{equation}
and for the variance, skewness, and excess kurtosis of $X$ in terms of those of
$X_1$ and $X_2$ \cite{statistics}
\begin{eqnarray}
\label{eqn:var_identity}
\mathrm{Var}[X]&=&\mathrm{Var}[X_1]+\mathrm{Var}[X_2],\\
\label{eqn:skew_identity}
\Tilde{\mu}_3[X]&=&\left(\dfrac{\mathrm{Var}[X_1]}{\mathrm{Var}[X_1]+\mathrm{Var}
[X_2]}\right)^{3/2}\Tilde{\mu}_3[X_1]+\left(\dfrac{\mathrm{Var}[X_2]}{\mathrm{Var}
[X_1]+\mathrm{Var}[X_2]}\right)^{3/2}\Tilde{\mu}_3[X_2],\\
\label{eqn:kurt_identity}
\mathcal{K}_{\rm ex}[X]&=&\left(\dfrac{\mathrm{Var}[X_1]}{\mathrm{Var}[X_1]+
\mathrm{Var}[X_2]}\right)^2\mathcal{K}_{\rm ex}[X_1]+\left(\dfrac{\mathrm{Var}
[X_2]}{\mathrm{Var}[X_1]+\mathrm{Var}[X_2]}\right)^2\mathcal{K}_{\rm ex}[X_2].
\end{eqnarray}
We proceed by first discussing the moments of $X_1$ and of $X_2$, and then we
will use these moments to extract closed-form expressions for the central
moments of $X$.

\subsubsection{Moments of \texorpdfstring{$X_1$}{TEXT} and \texorpdfstring{$X_2$}{TEXT}}
\label{sec:moments_X1_X2}

From Eq.~\eqref{eqn:p1_p2_def} it is clear that $X_1$ is a Gaussian random
variable with zero mean. Hence, its first two moments are given by
\begin{equation}
\langle X_1\rangle=0,\quad \langle X_1^2\rangle=\dfrac{k_BT}{\kappa},
\end{equation}
while its skewness and excess kurtosis are equal to zero, $\Tilde{\mu}_3
[X_1]=\mathcal{K}_{\rm ex}[X_1]=0$.

For $X_2$, instead, the central moments can be found via the identity
\begin{equation}
\label{eqn:central_moments_general}
\langle(X_2-\langle X_2\rangle)^n\rangle=\dfrac{1}{\mathrm{i}^n}\dfrac{d^n}{
dq^n}\left(\exp(-\mathrm{i}q\langle X_2\rangle)\hat{P}_2(q)\right)\biggr\rvert
_{q=0}=\dfrac{1}{\mathrm{i}^n}\dfrac{d^n}{dq^n}\left[\exp\left(\omega\tau\left(
I(q)-qI'(0)\right)\right)\right]\biggr\rvert_{q=0},
\end{equation}
where in the last expression we used that $\exp(-\mathrm{i}q\langle X_2\rangle)
=\exp(-(\omega\tau)qI'(0))$. It is shown in Appendix \ref{momentX2} that
\begin{equation}
\label{eqn:mean_X2}
\langle X_2\rangle=\omega\tau\langle Y\rangle_a,
\end{equation}
and that, for $n=2,3$,
\begin{equation}
\label{eqn:central_moments_X2_2_3}
\langle(X_2-\langle X_2\rangle)^n\rangle=\dfrac{\omega\tau}{n}\langle Y^n
\rangle_a,
\end{equation}
while for $n=4$ ,
\begin{equation}
\label{eqn:central_moments_X2_4}
\langle(X_2-\langle X_2\rangle)^4\rangle=\dfrac{1}{4}\omega\tau\langle Y^4
\rangle_a+\dfrac{3}{4}\omega^2\tau^2\langle Y^2\rangle_a^2.
\end{equation}
From the last two equations we can immediately get the variance, the skewness,
and the kurtosis of $X_2$,
\begin{equation}
\label{eqn:var_skew_kurt_X2}
\mathrm{Var}[X_2]=\dfrac{1}{2}\omega\tau\langle Y^2\rangle_a,\quad\Tilde{\mu}_3
[X_2]=\sqrt{\dfrac{8}{9\omega\tau}}\dfrac{\langle Y^3\rangle_a}{\sqrt{\langle
Y^2\rangle_a^3}},\quad\mathcal{K}_{\rm ex}[X_2]=\dfrac{1}{\omega\tau}\dfrac{
\langle Y^4\rangle_a}{\langle Y^2\rangle_a^2}.
\end{equation}

\subsubsection{Mean, variance, skewness, and excess kurtosis of \texorpdfstring{$X$}{TEXT}} 
\label{sec:mean_var_skew_kurt_X}

We can now obtain the mean, the variance, the skewness, and the excess kurtosis
of $X$ from expressions \eqref{eqn:moments_sum}, \eqref{eqn:var_identity},
\eqref{eqn:skew_identity}, \eqref{eqn:kurt_identity}, and the results of
Sec.~\ref{sec:moments_X1_X2}. Thus,
\begin{equation}
\label{eqn:X_mean}
\langle X\rangle=\omega\tau\langle Y\rangle_a,
\end{equation}
i.e., the average of the position is directly proportional to the average of
$\rho_a$ and to the rate of kicks $\omega$. The variance reads
\begin{equation}
\label{eqn:X_var}
\mathrm{Var}[X]=\dfrac{k_BT}{\kappa}+\dfrac{\omega\tau}{2}\langle Y^2\rangle_a.
\end{equation}
Hence, as expected from Eq.~\eqref{eqn:var_identity}, on the right hand side
a thermal and an active contribution appear. Interestingly, the active
contribution, which is basically the variance of $X_2$, scales as $\omega\tau$,
just as the mean. This linear dependence on $\omega\tau$ is an effect of the central limit theorem: both the mean and the variance must scale as the mean number of events in a characteristic time---which is indeed $\omega\tau$. 

The skewness of the process $X$ reads
\begin{equation}
\label{eq:skewness}
\Tilde{\mu}_3[X]=\sqrt{\frac{8}{9}}\omega\tau\left(\frac{2 k_BT}{k}+\omega\tau
\langle Y^2\rangle_a\right)^{-3/2}\langle Y^3\rangle_a.
\end{equation}
Interestingly, the distribution of $X$ is skewed if and only if $X_2$ is. Note
that having an amplitude distribution of kicks with zero skewness does not imply
that $X_2$ is not skewed as well. Finally, the excess kurtosis yields in the
form
\begin{equation}
\label{eq:exsskurtosis}
\mathcal{K}_{\rm ex}[X]=\omega\tau\langle Y^4\rangle_a\left(\frac{2k_BT}{
\kappa}+\omega\tau\langle Y^2\rangle_a\right)^{-2},
\end{equation}
which only depends on positive quantities, thus the distribution is always
leptokurtic. This result is in line with experimental observations for colloidal particles immersed in bacterial and engineered reservoirs~\cite{Krishnamurty16,Krishnamurty21}. 
Moreover, when considering $T$, $\kappa$, and the distribution $\rho_a(y)$ as fixed, we can find an optimal value of $\omega$ maximising the excess kurtosis,
\begin{equation}
\omega^*=\frac{2k_BT}{\kappa\tau\langle Y^2\rangle_a},
\end{equation}
returning the value $\mathcal{K}_{\rm ex}^{\star}[X]\equiv\mathcal{K}_{\rm ex}[X]\rvert_{\omega=\omega^*}$,
\begin{equation}
\mathcal{K}_{\rm ex}^{*}[X]=\frac{\kappa\langle Y^4\rangle_a}{8k_BT\langle
Y^2\rangle_a}.
\end{equation}
In the limit when $\omega\ll\tau$, when the kicks are rare, both the skewness
and the excess kurtosis vanish, and we retrieve the results of a Brownian
particle in a quadratic potential.
\subsection{Inferring the statistics of the underlying active kicks}
\label{sec:inf}
Let us now establish a link with realistic experimental scenarios.  Both the rate $\omega$ and the distribution $\rho_a(y)$, characterizing the PSN, are in general not directly measurable from experiments. On the other hand, the positions of the particles are directly measurable so it in general possible to have access to the distribution $P(x)$ and its stationary moments.
An interesting connection between the microscopic quantities $\omega$ and $\rho_a(y)$ and the distribution $P(x)$ was obtained by Kanazawa et al.~in \cite{kanazawaPRL}, where  the rate $\omega$ was related to the statistics of $X$ from the relation
\begin{equation}
    \omega = -\lim_{q\to \infty} \left[ q\dfrac{\partial \hat{P}(q)}{\partial q} + \dfrac{2k_BT}{\kappa} q^2 \right].\label{eq:kana}
\end{equation}
From the results derived in Sec.~\ref{sec:mean_var_skew_kurt_X} we add to the results of \cite{kanazawaPRL} the following relations which allow us to retrieve the moments of $Y$ from the moments of~$X$
\begin{align}
    \langle Y \rangle_a &= \dfrac{1}{\omega \tau}\langle X \rangle, \\
    \langle Y^2 \rangle_a &= \dfrac{2}{\omega \tau} \mathrm{Var}[X] - \dfrac{2k_BT}{\kappa \omega \tau},\\
    \langle Y^3 \rangle_a &= \dfrac{3}{\omega \tau} \Tilde{\mu}_3[X]\mathrm{Var}[X]^{3/2}, \\
    \langle Y^4 \rangle_a &= \dfrac{4}{\omega \tau} \mathcal{K}_{\rm ex}[X] \mathrm{Var}[X]^2.\label{eq:Y4inf}
\end{align}
Notably, Eqs.~(\ref{eq:kana}-\ref{eq:Y4inf}) provide a useful recipe for experimentalists to extract stationary statistics of the hidden stochastic process $Y_t$ which is generally not possible under the assumption of more complex dynamics.

\subsection{Effective temperature}
\label{effectiveT}
For a system described by an overdamped Langevin equation with active noise, one is often interested in defining an effective  temperature $T_{\mathrm{eff}}$~\cite{Maes14,Maes15,leticia}.  
In  close-to-equilibrium passive isothermal systems   an effective temperature can be defined by using  either the equipartition or fluctuation-response theorems. In active systems however, 
 mapping the non-equilibrium dynamics to an effective equilibrium dynamics  is fully reliable  whenever the stationary (or quasistatic) distribution is Gaussian  \cite{Viktor20,Viktor_Rahul_20,Zakine2017}. 
 
For non-Gaussian active dynamics, a possible definition of effective temperature is provided in terms the Onsanger regression principle~\cite{Onsanger1, Onsanger2}. We will briefly expose it here, but for a more detailed discussion we refer to \cite{crisantiJPA, Maes14, Maes15}. Let us consider that our system, described by \eqref{eqn:Langevin}, is at stationarity. Any observable $A_t$ has an average, that we denote with $\langle A_t \rangle$, and a correlation function defined as
\begin{equation}
    \label{eqn:Corr_def}
    C_A(t,s) = \langle A_t A_s \rangle.
\end{equation}
Since the system is at stationarity, the average of any observable does not depend on time, we can actually drop the dependence on $t$ and write $\langle A_t \rangle = \langle A \rangle$. At time $t=0$ we apply on the system a constant small force $\delta f$ that modifies the Langevin equation as follows
\begin{equation}
    \tau \dot{X}_t = - X_t + \sqrt{2 \tau \dfrac{k_BT}{\kappa}} \xi_t + \tau \eta_t + \dfrac{\delta f}{\kappa}.
\end{equation}
During the relaxation towards a new steady state, any observable $A$ has a time-dependent average $\langle A_t \rangle_{f} $ depending on the force $\delta f$. Onsanger regression principle states that the relaxation of the perturbed system towards the new steady state can be seen as a spontaneous equilibrium fluctuation. This translates into a simple equation \cite{crisantiJPA} connecting the correlation function $C_A(t,s)$ and the time-dependent response function (also known as susceptibility) $\chi_A(t)$ here defined as
\begin{equation}
    \label{eqn:chi_def}
    \chi_A(t) = \lim_{\delta f \to 0} \dfrac{\langle A_t \rangle_{f} - \langle A \rangle}{\delta f}.
\end{equation}
The Onsager relation for isothermal equilibrium systems  at temperature $T$ reads
\begin{equation}
    \chi_A(t) = \frac{ \left[ C_A(t,t) -C_A(t,0) \right]}{k_{\rm B}T}.
\end{equation}
Motivated by Onsager's relation we can introduce an  effective temperature by the relation
\begin{equation}
    \dfrac{1}{k_BT_{\rm eff}} = \dfrac{\chi_A(t)}{C_A(t,t) - C_A(t,0)}.\label{eq:efft1}
\end{equation}
When the time $t$ is large compared to the typical relaxation time of the system, in other words in the stationary limit $t \to \infty$, $A_t$ and $A_0$ become uncorrelated, i.e. $\langle A_t A_0 \rangle = \langle A \rangle^2$, thus we can rewrite Eq.~\eqref{eq:efft1} as
\begin{equation}
    \dfrac{1}{k_BT_{\rm eff}} = \dfrac{1}{\mathrm{Var}[A]} \lim_{\delta f \to 0} \dfrac{\langle A \rangle_{f} - \langle A \rangle}{\delta f}.\label{eq:efft2}
\end{equation}
 Specializing Eq.~\eqref{eq:efft1} to the position $A=X$ reads
\begin{equation}
    \dfrac{1}{k_BT_{\rm eff}} = \dfrac{1}{\mathrm{Var}[X]} \lim_{\delta f \to 0} \dfrac{\langle X \rangle_{\delta f} - \langle X \rangle_0}{\delta f} = \frac{1}{\text{Var}[X]}\lim_{t\to\infty}\chi_X(t).
\end{equation}
It is shown in appendix that the long-time susceptibility of the position reads $\lim_{t\to\infty}\chi_X(t)=1/\kappa$. Using this result together with the exact expression for the variance [Eq.\eqref{eqn:X_var}],  we get
\begin{equation}
    T_{\rm eff} = \dfrac{\kappa}{k_B} \mathrm{Var}[X] = T + \dfrac{\kappa \omega \tau}{2 k_B} \langle Y^2 \rangle_a.\label{eq:TeffFDTnew}
\end{equation}
In general the result of $T_{\rm eff}$ provided by the fluctuation-dissipation theory (FDT), is different from the one obtained from the equipartition theorem. Nevertheless, in the specific case in which PSN has zero mean, i.e. when $\langle Y \rangle_a=0$, according to equation \eqref{eqn:<X>} also $\langle X \rangle=0$, and so we have that ${\mathrm{Var}[X] = \langle X^2 \rangle}$. Thus in this situation, the equipartition theorem  leads to the same effective temperature than   the FDT. Therefore, the effective temperature could be  a useful tool to map the  second moment within the PSN into the second moment of an effective isothermal nonequilibrium (yet non-Gaussian) system.  We remark the presence of fat (leptokurtic) tails as a footprint of the nonequilibrium dynamics, as $P(X)\neq \exp(-V(x)/k_{\rm B}T_{\rm eff})$ is not Boltzmannian. Even though the variance can be captured by an effective Gaussian model at temperature $T_{\rm eff}$, it remains mandatory to employ a non-Gaussian description to capture higher-order moments such as skewness and kurtosis.
 %Nevertheless, a Gaussian approximation fails to capture the leptokurtyc tail of the position distribution.}
Below we now consider a particular case where kick amplitudes are drawn from the Gaussian distribution, as an application of this general discussion.

\section{Gaussian kicks}
\label{sec:Gaussian_kicks}
For illustrative purposes we now consider a particular form for the amplitude
distribution $\rho_a(y)$ of the active kicks amplitude, namely, the shifted
Gaussian distribution
\begin{equation}
\label{eqn:rho_Gaussian}
\rho_a(y)=\dfrac{1}{Z}\exp\left(-\dfrac{\beta_{a}\gamma^2}{2m_a}\left[y-
\dfrac{\mu}{\gamma}\right]^2\right).
\end{equation}
This distribution \eqref{eqn:rho_Gaussian} is analogous to the equilibrium
distribution of a free particle of mass $m_a$ with average momentum $\mu$
and immersed in a thermal bath with temperature
\begin{equation}
T_a=\frac{1}{k_B\beta_a}.
\end{equation}
Here $T_a$ is an effective temperature as it has the dimensions of temperature
but it is a parameter that is not constrained by any fluctuation-dissipation
relations enforcing a specific relation with the friction coefficient $\gamma$
of the particle. Similarly, $m_a$ can be understood as an effective mass of the
components of the active bath that kick the Brownian particle. Furthermore, the
quantity $Z=\sqrt{2\pi m_a/\beta_a\gamma^2}$ in Eq.~\eqref{eqn:rho_Gaussian} is
a normalisation factor enforcing normalisation, $\int_{-\infty}^{\infty}\rho_a
(y)dy=1$, and can thus be viewed as an effective partition. For analytical ease
it is useful to introduce the quantities
\begin{equation}
\tau_a\equiv\frac{m_a}{\gamma},\quad\lambda\equiv\frac{\mu}{\gamma}.
\end{equation}
Here, $\tau_a$ is an effective momentum relaxation time scale, and $\lambda$ is
a length scale given by the mean value of the Gaussian kick amplitude
distribution.

We now use the expressions derived in Subsection \ref{sec:moments} to inspect
the stationary moments of the particle. Specialising Eqs.~\eqref{eqn:X_mean}
and \eqref{eqn:X_var} for the Gaussian PDF \eqref{eqn:rho_Gaussian} we find
that the mean and the variance of the position in this case become
\begin{equation}
\label{eqn:mwan_var}
\langle X\rangle=\omega\tau\lambda,\quad\mathrm{Var}[X]=\dfrac{k_BT}{\kappa}+
\dfrac{\omega\tau}{2}\left(k_BT_a\dfrac{\tau_a}{\gamma}+\lambda^2\right).
\end{equation}
These formulas have a simple interpretation: the average position is proportional
to the average amplitude $\mu$ of the kicks and to their occurrence rate. At the
same time, it is inversely proportional to the stiffness of the potential: the
stiffer the potential is the more the particle is forced to stay close to the
origin. The variance of the position is given by the sum of two contributions:
$k_BT/\kappa$ associated with the thermal bath, and the contribution due to the
active kicks. The contribution of the second term can be manipulated externally
by changing the bacterial activity, e.g., by tuning the rate $\omega$ of active
kicks, the mean amplitude $\mu$, and variance, which can be controlled via $m_a$
and/or $T_a$. In practice, a bacterial suspension will be associated with given
values of $\beta_a$ and $m_a$, that may depend on the bacterial metabolic
activity, division rate, etc. Conversely, $\mu$ may be imposed externally by,
e.g., pressure-based microfluidic flow-control systems in fluid chambers with
periodic boundary conditions. We note that it is particularly instructive to
consider the  most common experimental setting corresponding to the case of
closed boundary conditions \cite{Krishnamurty16,Krishnamurty21}, $\lambda=0$,
for which we still get an enhancement of the variance with respect to the
thermal reference value $k_BT/\kappa$.

{As we have shown in Eq.~\eqref{eq:TeffFDTnew}, the usage of equipartition theorem in recent experimental work in \cite{Krishnamurty16,Krishnamurty21} is well-suited
for the case $\mu=0$, }
\begin{equation}
T_{\rm eff}=\frac{\kappa\langle X^2\rangle}{k_B}.
\label{eq:TeffAjay}
\end{equation}
For an equilibrium Langevin dynamics in the external potential $U(x)=\kappa x^2
/2$, $T_{\rm eff}$ would be the temperature needed to attain the value $\langle
X^2\rangle$ of the second moment observed in the PSN active-bath model. For the
PSN active noise with symmetric Gaussian kick amplitudes, we get via
Eq.~\eqref{eqn:mwan_var} for $\lambda=0$ that
\begin{eqnarray}
\label{eq:TeffCBC}
T_{\rm eff}=T+\left(\dfrac{\omega\tau_a}{2}\right)T_a.
\end{eqnarray}
As expected, the PSN active noise leads to an effective temperature, that is
always larger or equal than the bath temperature, i.e., $T_{\rm eff}\geq T$.
Equation \eqref{eq:TeffCBC} implies that the effective temperature defined
by \eqref{eq:TeffAjay} is sensitive to two properties of the active bath,
the rate of kicks $\omega$, and the variance $\sim\tau_a T_a\propto T_am_a$
of the kick amplitudes. {%We note that when the distributions are Gaussian (or have finite width) the effective temperature could be a useful quantity.  We calculate the effective temperature because of its experimental relevance, as measured in \cite{Krishnamurty16,Krishnamurty21}. 
For the case of colloidal heat engines in bacterial reservoirs  the effective temperatures derived through Eq.~\eqref{eq:TeffAjay} were found to be up to one order of magnitude larger than the room temperature. For example in \cite{Krishnamurty16} the bath temperature during the isothermal steps was $T\sim 300K$ but the active temperature was found to be $T_{\rm eff}\sim1000K$. Such enhancement of the effective temperature was linked to the efficiency enhancement in theoretical models of  active heat engines~\cite{Marathe19, Marathe22, Ruben23, Viktor20, Viktor_Rahul_20} where exact non-Gaussian statistics and fluctuation response were not tackled.}

We also compute the skewness of the distribution from Eq.~\eqref{eq:skewness},
\begin{eqnarray}
\Tilde{\mu}_3[X]=\dfrac{\omega\tau\lambda}{3}\left(3k_BT_a\dfrac{\tau_a}{\gamma}
+\lambda^2\right)\left(\dfrac{k_BT}{\kappa}+\dfrac{\omega\tau}{2}\left(k_BT_a
\dfrac{\tau_a}{\gamma}+\lambda^2\right)\right)^{-3/2}.
\end{eqnarray}
This result shows that in the specific case of symmetric or extremely rare kicks,
i.e., small $\mu$ and $\omega$, the skewness vanishes, as it should.

Let us conclude this Subsection with the exact analytical expression for the
excess kurtosis [see Eq.~\eqref{eq:exsskurtosis}], which yields in the form
\begin{eqnarray}
\label{eq:exsskurtosis_Gaussiankick}
\mathcal{K}_{\rm ex}[X]=\dfrac{3\omega\tau}{4}\left(\left(k_BT_a\dfrac{\tau_a}{
\gamma}\right)^2+\frac{\lambda^2}{3}\left(6k_BT_a\dfrac{\tau_a}{\gamma}+\lambda^2
\right)\right)\left(\dfrac{k_BT}{\kappa}+\dfrac{\omega\tau}{2}\left(k_BT_a
\dfrac{\tau_a}{\gamma}+\lambda^2\right)\right)^{-2}.
\end{eqnarray}
Equation~\eqref{eq:exsskurtosis_Gaussiankick} confirms that for any parameter
value $\mathcal{K}_{\rm ex}[X]\geq 0$ is always positive, indicating that the stationary distribution of the position is always leptokurtic in the case of Gaussian kicks. 
% and increases with rising activity, up to a certain range.
The detailed analysis of $\mathcal{K}_{\rm
ex}[X]$ is presented in Subsection \ref{subsec:excsskurtosis} for various
parameter regimes.

\subsection{Results for \texorpdfstring{$\mathcal{K}_{\rm ex}$}{TEXT} and
deviation from the Gaussian distribution to non-Gaussianity and re-entry
to Gaussianity}
\label{subsec:excsskurtosis}

\begin{figure}
\centering
\includegraphics[width=8cm,height=6cm,angle=0]{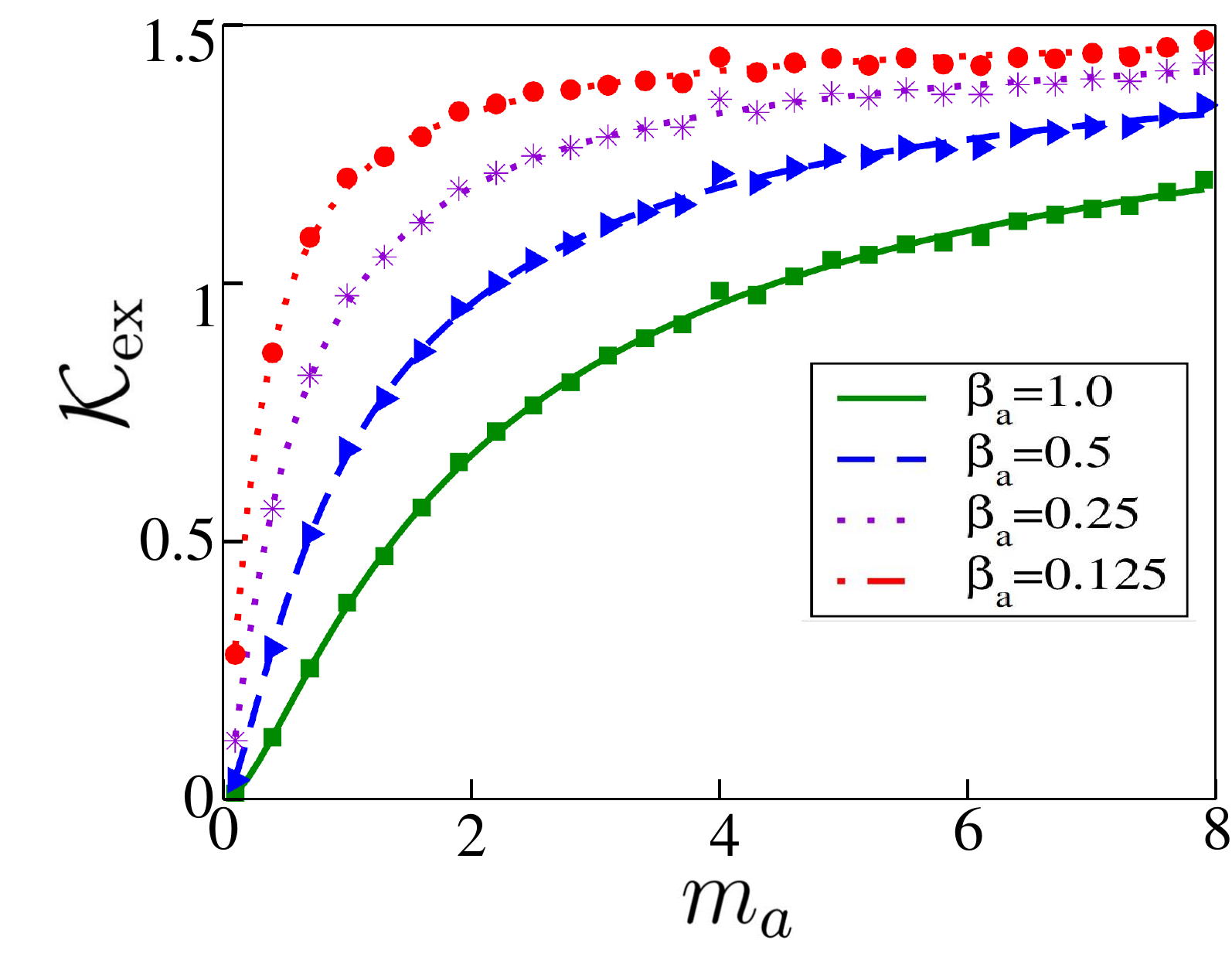}
\includegraphics[width=8cm,height=6cm,angle=0]{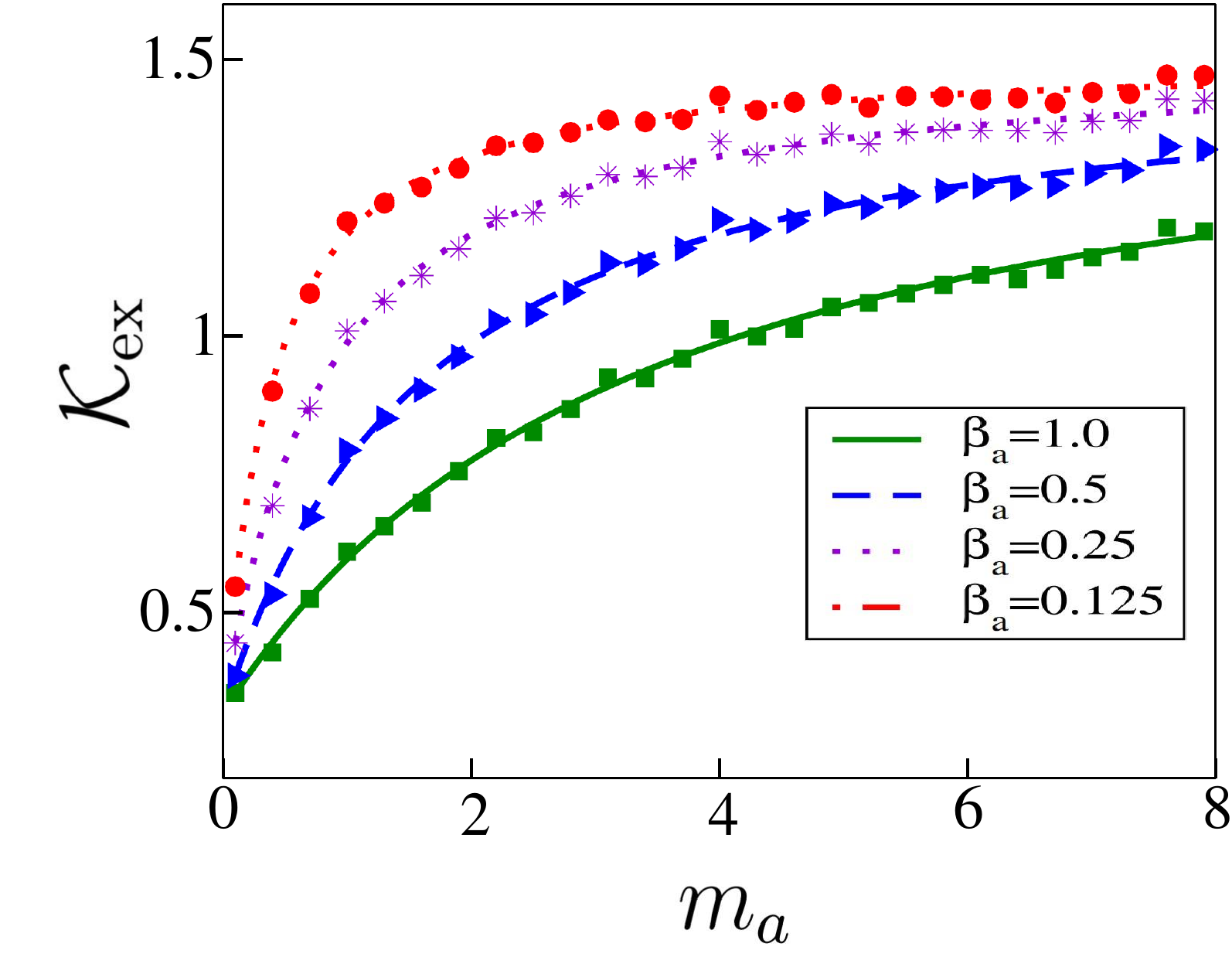}
\caption{Excess kurtosis $\mathcal{K}_{\rm ex}$ as a function of the mass
$m_a$ of the active particle for different values of $\beta_{a}$, for two different values of the active bath's mean kick amplitude: {  $\mu
=0$ (left panel), and $\mu=2.0$ (right panel).
Other simulation  parameters were $\omega=2.0$, $T=0.5$, $\kappa=2.0$, and $\gamma=2.0$, total
number of simulations  $10^5$, and simulation time step $dt=10^{-4}$. In both panels, symbols are obtained from numerical simulations while lines are theoretical predictions given by Eq.~\eqref{eq:exckurtheo}.}}
\label{Kurt_I}
\end{figure}

This Section briefly discusses and concludes the analytical and numerical
results of $\mathcal{K}_{\rm ex}$ for the particular Gaussian form for the
amplitude distribution $\rho_a(y)$ given by Eq.~\eqref{eqn:rho_Gaussian}. The behaviour of the Brownian particle in the presence
of active PSN exhibits non-Gaussian dynamics. In the steady state, the excess
kurtosis is quantified by Eq.~\eqref{eq:exsskurtosis_Gaussiankick}. In
Fig.~\ref{Kurt_I} we illustrate the relationship between $\mathcal{K}_{\rm
ex}$ and the mass $m_a$ of the active particle across various values of
$\beta_a$. Here, we examine two scenarios. In the left panel in Fig.~\ref{Kurt_I}, we explore the
behaviour of the excess kurtosis in a setup where the particle is optically
trapped within a closed container filled with bacteria. In this configuration,
the active self-propelled bacteria exert net drift ($\mu=0$) and
move randomly inside the fluid container. The expression of the excess
kurtosis in this case reads,
\begin{equation}
\mathcal{K}_{\rm ex}[X]=\frac{3\omega\tau_a^2}{\tau}\left(\omega\tau_a+2
\dfrac{T}{T_a}\right)^{-2}.
\label{eq:exckurtheo}
\end{equation}
We further compare these with the behaviour of the kurtosis in the limits of
$m_a$. It is evident that the mass of the bacteria plays a pivotal role in
determining the system's dynamics. When $m_a$ approaches zero, $\tau_a$
also tends to zero, resulting in the convergence to zero of the excess
kurtosis $\mathcal{K}_{\rm ex}$. Conversely, for large values of $m_a$ and
as $\tau_a$ approaches very large values, the excess kurtosis converges
to $3/\omega\tau$. 

In the right panel in Fig.~\ref{Kurt_I}, we
consider the scenario of an optically trapped particle within a bacterial
bath under a uniform fluid flow, as illustrated in Fig.~\ref{schematic_diagram}.
At large $m_a$ (i.e. in the limit of $\tau_a\to\infty$) the value of the excess
kurtosis increases and eventually saturates at the value $3/\omega\tau$
for fixed values of $T_a$ and $\omega$. Furthermore, when we increase the
active temperature $T_a$ for given $m_a$ and $\omega$, the non-Gaussianity of
the system becomes more pronounced.% In the limit of $T_{a}\to\infty$, $\mathcal{K}_{\rm ex}$  converges to $3/\omega\tau$ for both scenarios.

The preceding analysis thus reveals that the excess kurtosis reaches its asymptotic large $m_a$ limit that is independent on the external torque $\mu$. Furthermore, the active temperature $T_a$ controls the sharpness of the transition between Gaussian ($\mathcal{K}_{\rm ex}=0$) to non-Gaussian ($\mathcal{K}_{\rm ex}>0$) behaviour; the larger $T_a$ the steeper is the transition from the minimum to the maximum value of the excess kurtosis as a function of~$m_a$.

\begin{figure}[h]
\centering
\includegraphics[width=8cm,height=6cm,angle=0]{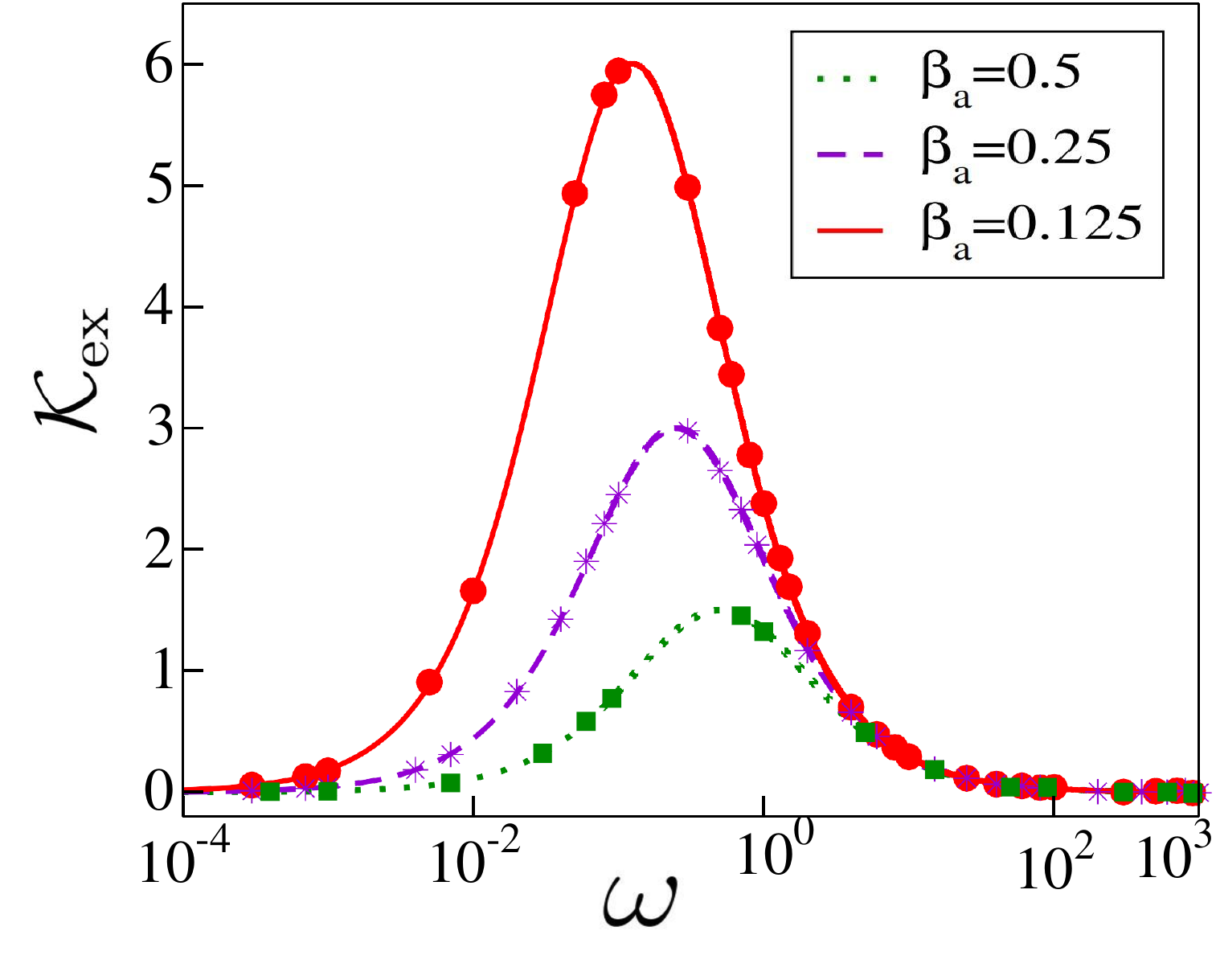}
\includegraphics[width=8cm,height=6cm,angle=0]{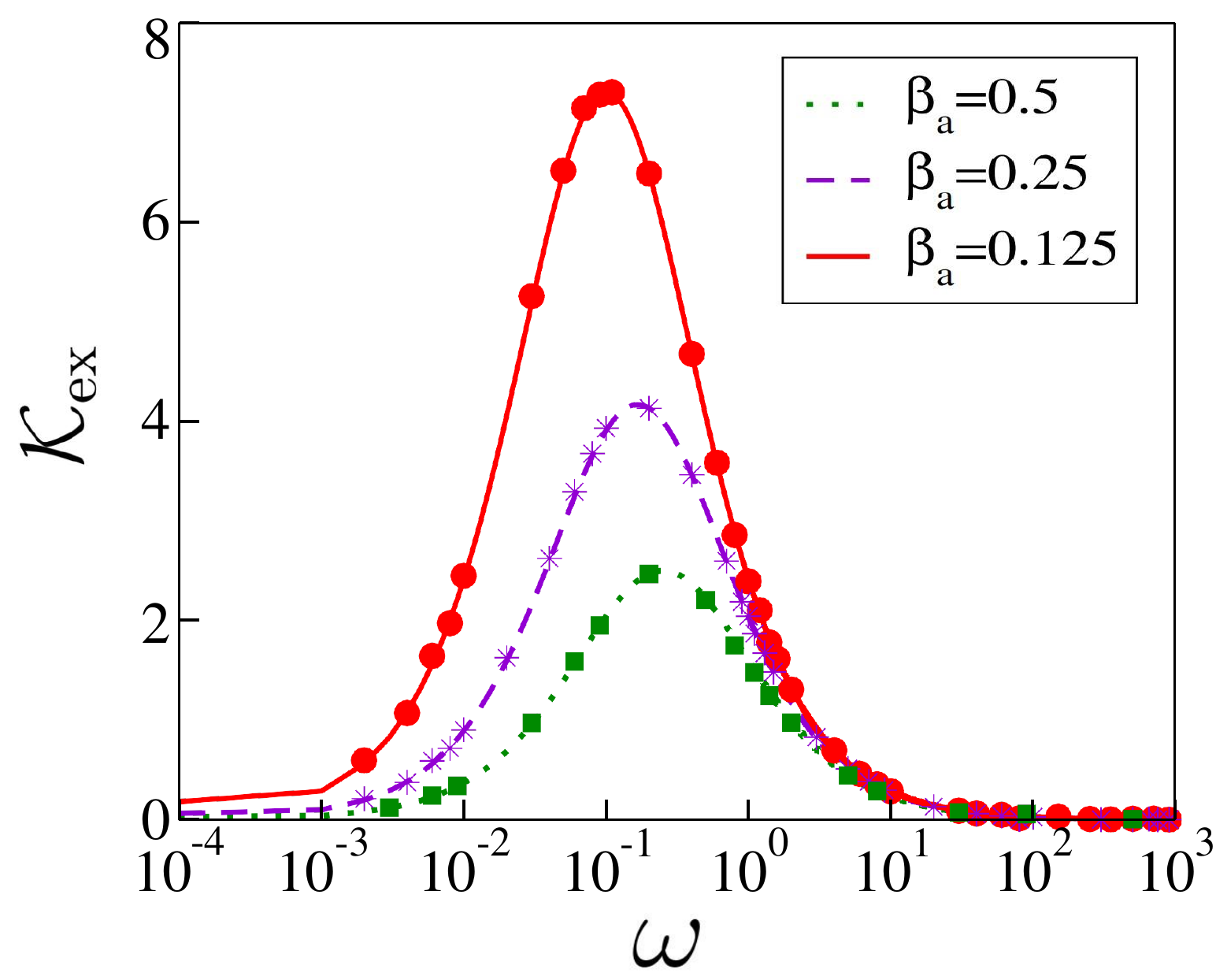}
\caption{ (Left panel) Excess kurtosis $\mathcal{K}_{\rm ex}$ as a function of the kick rate $\omega$ for $\mu=0.0$ for different  values of $\beta_{a}$. (Right panel) Excess kurtosis $\mathcal{K}_{\rm ex}$ as a function of the kick rate $\omega$ for $\mu=2.0$. Here other Parameters used are  $m_a=2.0$, $T=0.5$, $\kappa=2.0$,  $\gamma=2.0$,  number of simulations  $10^5$ and simulation time step $dt=10^{-4}$. { In both panels, symbols are obtained from numerical simulations while lines are theoretical predictions given by Eq.~\eqref{eq:exckurtheo}.}}  
\label{Kurt_II}
\end{figure}

Next we analyze in Fig.~\ref{Kurt_II} how the excess kurtosis $\mathcal{K}_{\rm ex}$
depends on the kick rate $\omega$. 
In the left panel in Fig.~\ref{Kurt_II} we observe that the degree of non-Gaussianity as measured by $\mathcal{K}_{\rm ex}$ exhibits a non-monotonic dependency with $\omega$. %When $\omega$ is small, interactionsof the Brownian particle with bacteria are rare, resulting in a predominantly Gaussian behaviour. 
In the limit of $\omega$ small, the system has no active
particles and behaves as a thermal system, resulting in a predominantly Gaussian behaviour. As we increase $\omega$, the number
of kicks between the Brownian particle and the active particles becomes
significant, effecting a substantial enhancement of the system dynamics,
pushing it out of equilibrium and increasing the non-Gaussian behaviour.
Interestingly, however, in the limit of $\omega$ large, the Brownian particle
is kicked by the active particles even more frequently and in a more random
fashion. Consequently, the system tends towards Gaussian white noise with
finite intensity, restoring Gaussianity. %In summary, $\omega\to0$ and $\omega\to\infty$ represent the two passive limits.
This process illustrates
the transition from Gaussian to non-Gaussian behaviour and the subsequent
return to Gaussianity, known as re-entry into the Gaussian regime. This
signifies the crossover from passive to active and back to (effectively)
passive behaviour.

Furthermore, by keeping $T$, $\kappa$, and the distribution $\rho_a(y)$ fixed,
it is possible to identify an optimal value of $\omega$ that maximises the
excess kurtosis. This optimal value for $\omega$ reads
\begin{equation}
%\omega^*=\frac{2\beta_ak_BT\gamma}{m_a+\beta_a\mu^2},
\omega^*=\dfrac{2 k_{B}T\beta_{a}}{\tau_a+\lambda^2\beta_a\gamma},
\label{eq:wmax}
\end{equation}
corresponding to the maximum value of $ \mathcal{K}_{\rm ex}[X]$,
\begin{equation}
%\mathcal{K}_{\rm ex}^{*}[X]=\frac{(3m_a^2+6m_a\beta_a\mu^2+\beta_a^2\mu^4)}
%{8\tau\beta_ak_BT\gamma(m_a+\beta_a\mu^2)}.
%\mathcal{K}_{\rm ex}^{*}[X]=\left(\dfrac{3\tau_a^2}{\beta_a}+6\tau_a\lambda^2\beta_a\gamma+\beta_a\gamma^2\lambda^4\right)\left(8\tau k_{B}T\left(\tau_a+\beta_a\gamma\lambda^2\right)\right)^{-1}.
\mathcal{K}_{\rm ex}^{*}[X]=\left(\dfrac{\dfrac{3\tau_a^2}{\beta_a}+6\tau_a\lambda^2\beta_a\gamma+\beta_a\gamma^2\lambda^4}{8\tau k_{B}T\left(\tau_a+\beta_a\gamma\lambda^2\right)}\right).
\label{exkurt_max}
\end{equation}

\begin{figure}[h]
\centering
\includegraphics[width=18cm]{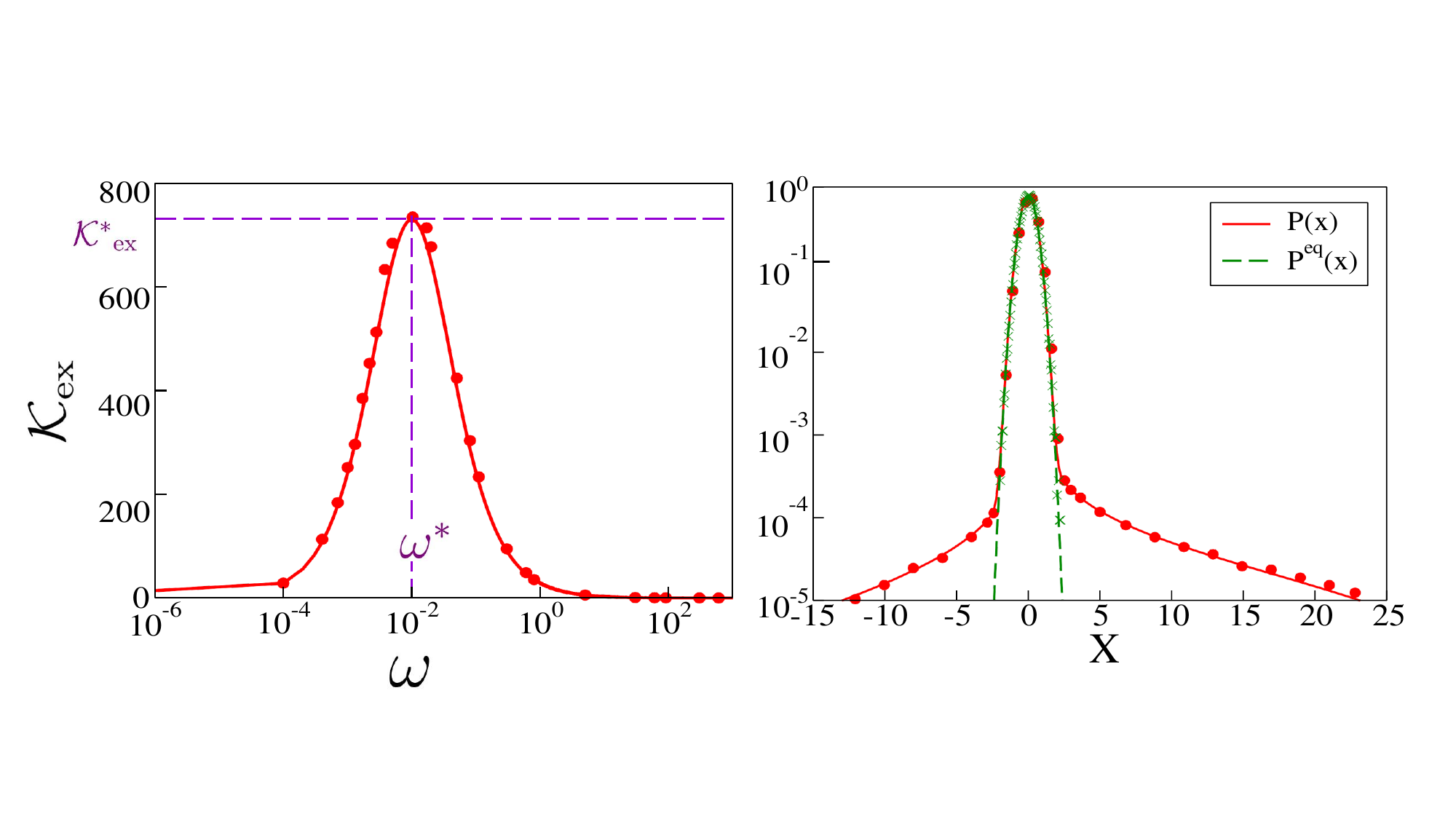}
\caption{{(Left panel) Excess kurtosis $\mathcal{K}_{\rm ex}$ as a function of
the kick rate $\omega$ obtained from numerical simulations (red circles) and compared with the theoretical prediction (red solid line, Eq.~\eqref{eq:exckurtheo}). The vertical dashed purple line is set at the optimal value of the kick rate $\omega^{*}$ [Eq.~\eqref{eq:wmax}] maximizing the excess kurtosis at a value $\mathcal{K}_{\rm ex}^{\star}$ [Eq.~\eqref{exkurt_max}]. (Right panel) Probability density for the position of the Brownian particle obtained from numerical obtained from numerical simulations (red circles). The lines are given by the analytical solution of the stationary density ($P(x)$, red solid line) and the (Gaussian) equilibrium distribution in the absence of kicks  ($P^{\rm eq}(x)$, green dashed line). Other simulation parameters were $m_a=2.0$, $T=0.5$, $\kappa=
2.0$, $\gamma=2.0$, $\mu=0.0$,  number of simulations  $10^5$, and  simulation time
step $dt=10^{-4}$.}}
\label{dist_I}
\end{figure}

In Figure \ref{dist_I}, the left panel illustrates the relationship between
the excess kurtosis $\mathcal{K}_{\rm ex}$ and the rate $\omega$ of kicks
for a specific value of $T_a$. Here Eq.~\ref{exkurt_max} is used to identify
the maximum value $\mathcal{K^{*}}_{\rm ex}$ and the corresponding optimal
rate $\omega^*$. In the right panel, we use the value of $\omega^*$ to
examine the displacement PDF of the Brownian particle and compare it with
the Gaussian PDF. The plot distinctly illustrates that the active distribution
exhibits   tails that are significantly wider than in comparison to the shorter tails of the
Gaussian distribution associated with the equilibrium state $\omega=0$.

\section{Conclusion and discussion}
\label{sec:conclusion}

We presented a combined analytical and numerical approach to compute higher order moments and the PDF for a colloidal particle in a thermal bath which experiences additional Poisson-shot noise in an active environment for the case of a harmonic trapping. Our formulation includes a general amplitude
PDF of the kicks due to the active particles. The latter are considered for both cases with and without a net drift. In the non-equilibrium steady state we quantify the skewness and non-Gaussianity of the emerging PDF and discuss the effective temperature of the system. Concretely for a Gaussian amplitude PDF of the shot noise we obtain exact results for the excess kurtosis and demonstrate that the stationary PDF is leptokurtic, with a narrow central region and heavy tails.

{ Our model and analysis provides a flexible platform to describe one-dimensional experimental systems that are characterized by nonequilibrium stationary states with fat tails. A key avantage with respect to other solvable models, such as the AOUP, is the fact that the stationary state of our model is in general non-Gaussian. Moreover, unlike many ad-hoc active-matter accounts, our model reconciles  the fluctuation-dissipation theorem and the equipartition theorem for any active kick distribution with zero mean. This makes our description  of singular interest to experimentalists who will likely benefit from our inference method developed in Sec.~\ref{sec:inf}. We expect our approach to impact ongoing research on stochastic thermodynamics of active systems, for which we outline some preliminary ideas below.}

The analytical approach developed here may find applications in the stochastic thermodynamics of active matter. We here add some remarks on some open
questions regarding irreversibility and dissipation associated with the dynamics of the non-equilibrium stationary state induced by the active
PSN. As noted, since the foundations of stochastic thermodynamics,
quantifying the asymmetry under time reversal of a stationary time series,
provides means to estimate the underlying entropy production of the physical
mechanism generating the time series \cite{Roldan2010}. Within this context,
it was shown that the rate of irreversibility measured by the Kullback-Leibler
(KL) divergence rate
\begin{equation}
\label{eq:KLD}
\sigma=\lim_{t\to\infty}\frac{1}{t}\int\mathcal{D}x_{[0,t]}P(x_{[0,t]})\ln
\frac{P(x_{[0,t]})}{P(\Theta_tx_{[0,t]})}\geq0,
\end{equation}
provides a lower bound to the steady-state rate of entropy production.
Equation \eqref{eq:KLD} can be understood as the KL divergence rate between
the probability of trajectories and their time reversal. In other words,
$P(x_{[0,t]})=P(X_0=x_0,\ldots X_t=x_t)$ is the probability to observe the
sequence $x_0,\ldots,x_t$ in the time interval $[0,t]$, and $P(\Theta_t
x_{[0,t]})=P(X_0=x_t,\ldots X_t=x_0)$ the probability to observe the
time-reversed sequence $x_t,\ldots,x_0$ in the same time interval $[0,t]$,
where both $P$ and $P(\Theta_t)$ are evaluated at the stationary state. Our
active-matter model given by Eq.~\eqref{eqn:Langevin} displays time
irreversibility in the presence of PSN. Indeed by inspection of the time
series shown in the right panel of Fig.~\ref{schematic_diagram}) one finds
that rapid changes in the particle position due to active kicks are often
accompanied by slow relaxations---a dynamics being time irreversible even
for the case of symmetric PSN with vanishing net velocity $\mu=0$. Thus, in
general one has $\sigma>0$ as a signature of irreversibility whenever $\omega
>0$. Conversely, since the dynamics is bounded by the confining potential, one
can show that the particle energy $U_t=(1/2)\kappa X_t^2$, which fluctuates
over time, is conserved on average, leading to a vanishing heat dissipation
rate. This leads us to conclude that traditional probes of irreversibility
such as heat dissipation are not sufficient to characterise the non-equilibrium
features of the PSN; instead, one should take an information-theoretical
approach, e.g., by evaluating $\sigma$ in Eq.~\eqref{eq:KLD}, or using
cross-correlation asymmetries as discussed below.

\begin{figure}
\centering
\includegraphics[width=16.0cm]{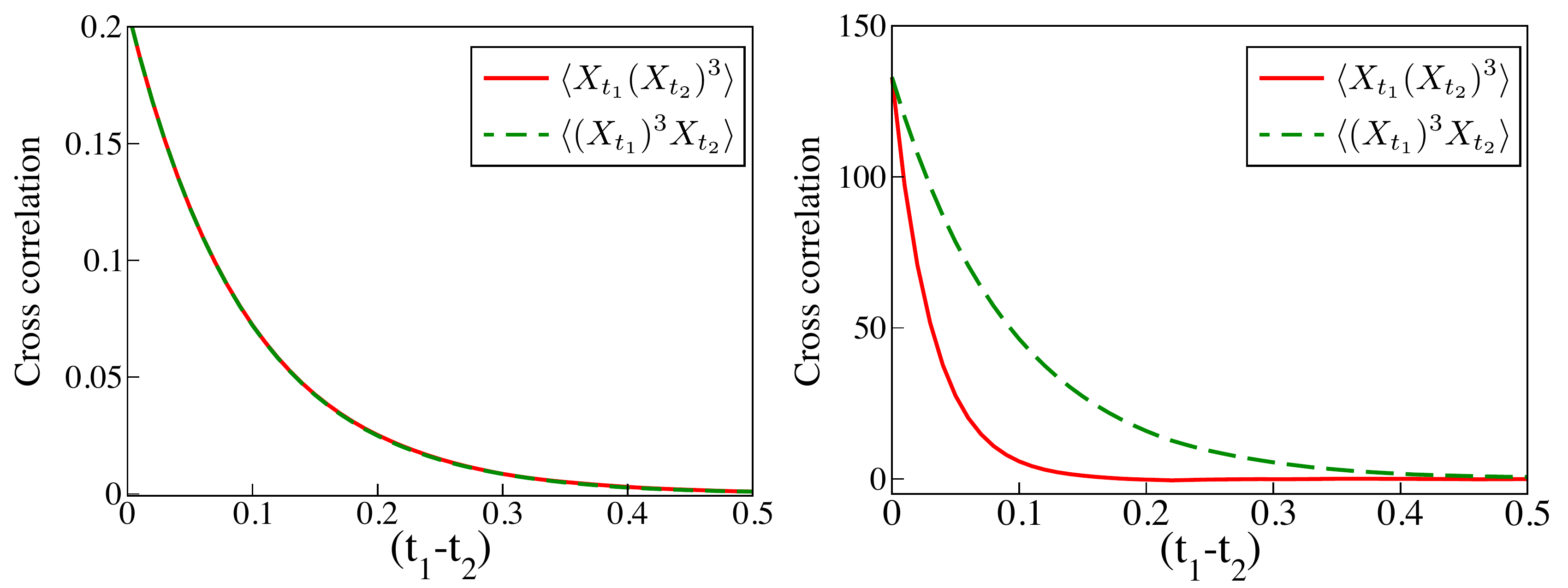}
\caption{Cross-correlation asymmetry in the presence of active Poissonian shot
noise (PSN) with amplitudes drawn from the Gaussian distribution
\eqref{eqn:rho_Gaussian}. Left panel: Comparison between the forward and
backward $X,X^3$ cross-correlation functions \eqref{eq:xcross} for an
equilibrium dynamics given by Eq.~\eqref{eqn:Langevin} without active
PSN (i.e., vanishing rate of kicks $\omega=0$). Right panel: Comparison between the forward and
backward $X,X^3$ cross-correlation functions \eqref{eq:xcross} in the presence of PSN with  rate of kicks 
$\omega=10^{-2}$. Other simulation parameters  are $T=0.5$, $\kappa=2.0$, $\gamma=0.2$,
$m_a=2.0$, $\mu=0.0$, and $T_a=8.0$, number of simulations  $10^5$, and
simulation time  step $dt=10^{-4}$.}
\label{fig:irreversibility}
\end{figure}

Evaluating the irreversibility rate \eqref{eq:KLD} for our model is not an
easy task, both analytically and from numerical estimates. We here discuss
some alternatives to quantify the degree of time irreversibility. First we
note that because the dynamics is one-dimensional and with open boundary
conditions, the stationary probability current vanishes. This implies that
$P(x_{t_1},x_{t_2})=P(x_{t_2},x_{t_1})$; in other words, one cannot detect
irreversibility from the autocorrelation function of the position. In general,
however, $P(x_{t_1},x_{t_2},x_{t_3})\neq P(x_{t_3},x_{t_2},x_{t_1})$ even in
the absence of a net current in $X$ \cite{Roldan2010}, which reveals that the
$(n\geq3)$th time correlators of the particle position $\langle X_{t_1}\ldots
X_{t_n}\rangle$ could be used to characterise the irreversibility of our model.
Following Steinberg \cite{Steinberg}, an alternative approach is to consider
two-time correlators including non-linear functions of the observables. For
example, let us consider the following two-time cross-correlators of the
position and of its third power,
\begin{equation}
\label{eq:xcross}
C_{t_1,t_2}^{X,X^3}\equiv\langle X_{t_1}(X_{t_2})^3\rangle,\quad C_{t_1,
t_2}^{X^3,X}\equiv\langle(X_{t_1})^3 X_{t_2}\rangle.
\end{equation}
Figure \ref{fig:irreversibility} shows numerical estimates of the correlators
$C_{t_1,t_2}^{X,X^3}$ (red solid line) and $C_{t_1,t_2}^{X^3,X}$ (green dashed
line) for the case of symmetric Gaussian kicks ($\mu=0$) with zero kick rate
$\omega=0$ (left panel) and non-zero kick rate $\omega>0$ (right panel). Our
analysis reveals an asymmetry between the $X,X^3$ cross-correlators $C_{t_1,
t_2}^{X,X^3}\neq C_{t_1,t_2}^{X^3,X}$ in the presence of active kicks for all
values of $t_1$ and $t_2$ that we explored, and vice versa a symmetry $C_{t_1,
t_2}^{X,X^3}=C_{t_1,t_2}^{X^3,X}$ for all $t_1,t_2$ in the absence of active
noise. Our numerical result reinforces recent insights from stochastic
thermodynamics which have unveiled the difference of cross-correlators, e.g.,
$\left|C_{t_1,t_2}^{X,X^3}-C_{t_1,t_2}^{X^3,X}\right|$, as probes of the
degree of  non-equilibrium \cite{Ohga,Shiraishi,Vu,liang2023thermodynamic,
Oberreiter}. Such results have provided lower bounds to the rate of
irreversibility $\sigma$ of Markovian systems that are directly proportional
to the asymmetry of cross correlators. It will be interesting in the future
to relate our findings to these novel approaches to tackle irreversibility
and dissipation through non-trivial cross-correlation structures.

We finally remark that while the harmonic oscillator is the most fundamental
model in statistical physics, cases of anharmonic external potentials should
also be considered in the presence of fluctuating forces different from white
Gaussian noise. Thus, processes with Gaussian yet long-range dependent noises
exhibit non-Boltzmannian stationary PDFs in the presence of steeper than
harmonic potentials or do not possess a stationary state in shallower than
harmonic potentials \cite{tobias,tobias1,vojta}. It will thus be interesting
to see how particles driven by both thermal noise and PSN perform in anharmonic
potentials.

\medskip
\textbf{Acknowledgements} \par %delete if not applicable))
We thank  Kiyoshi Kanazawa and Dario Lucente for useful discussions. Rah. M. gratefully acknowledges Science and Engineering Research
Board (SERB), India for financial support through the MATRICS Grant
(No. MTR/2020/000349). Ri. M. acknowledges Sandwich Training
Educational Programme (STEP) by Abdus Salam International Center for
Theoretical Physics (ICTP), Trieste, Italy.
Ra. M. acknowledges the German Science Foundation (DFG, grant ME 1535/12-1)
for financial support.
\'E.~R. acknowledges financial support from PNRR MUR project PE0000023-NQSTI.

\medskip

% Use the following code if you wish to generate your bibliography with BibTeX;
% replace the string "MSP-template" below with the name(s) of
% the BibTeX data base(s) you want to use.
% The resulting bibliography-output (the content of the .bbl file)
% must be pasted back into this file before submission.
% Please also include your BibTeX data base file(s) in your submission
% so that we can re-run BibTeX if necessary.
%
%\bibliographystyle{MSP}
%\bibliography{MSP-template}

\begin{thebibliography}{99}

\bibitem{einstein} A. Einstein, Ann. Phys. (Leipzig) \textbf{322}, 549 (1905).

\bibitem{smolu} M. von Smoluchowski, Ann. Phys. (Leipzig) \textbf{21}, 756 (1906).

\bibitem{langevin} P. Langevin, C. R. Acad. Sci. (Paris) \textbf{146}, 530 (1908).

\bibitem{levy} P. L{\'e}vy, Processus stochastiques et mouvement brownien
(Gauthier-Villars, Paris, 1948).

\bibitem{zwanzig} R. Zwanzig, Nonequilibrium statistical mechanics
(Oxford University Press, Oxford UK, 2001).

\bibitem{brenig} W. Brenig, Statistical theory of heat: Nonequilibrium
phenomena (Springer, Berlin, 1989).

\bibitem{toda} R. Kubo, M. Toda, and N. Hashitsume, Statistical physics II:
Nonequilibrium statistical mechanics (Springer, Berlin, 1985).

\bibitem{vankampen} N. G. van Kampen, Stochastic processes in physics and
chemistry (North-Holland, Amsterdam, 1981).

\bibitem{landau} L. D. Landau and E. M. Lifshitz, Landau and Lifshitz Course
of Theoretical Physics 5: Statistical Physics Part 1 (Butterworth-Heinemann,
Oxford UK, 1980).

\bibitem{sekimoto98} K. Sekimoto, Prog. Theoret. Phys. \textbf{130},  17-27 (1998).

\bibitem{Seifert12} U. Seifert, Rep. Prog. Phys. \textbf{75}, 126001 (2012).

\bibitem{Vicsek12} T. Vicsek and A. Zafeiris, Phys. Rep. \textbf{517}, 71 (2012).

\bibitem{Ramaswamy10} S. Ramaswamy, Ann. Rev. Cond. Mat. Phys. \textbf{1}, 323
(2010).

\bibitem{Volpe16} C. Bechinger, R. D. Leonardo, H. L{\"o}wen, C. Reichhardt, G.
Volpe and G. Volpe, Rev. Mod. Phys. \textbf{88}, 045006 (2016). 

\bibitem{Krishnamurty16} S. Krishnamurthy, S. Ghosh, D. Chatterji, R. Ganapathy,
and A. K. Sood, Nat. Phys. \textbf{12}, 1134 (2016).

\bibitem{leticia}{ D. Loi, S. Mossa, and L. F. Cugliandolo, Phys. Rev. E \textbf{77} (5), 051111 (2008).}

\bibitem{Maes14} C. Maes, J. Stat. Phys. \textbf{154}, 705-722 (2014).

\bibitem{Maes15} C. Maes and S. Steffenoni, Phys. Rev. E, {\bf 91}, 022128 (2015).

\bibitem{Cates} E. Fodor,  et al., Phys. Rev. Lett. {\bf 117}, 038103 (2016).

\bibitem{Genesotto18} F. S. Gnesotto, F. Mura, J. Gladrow, and C. P. Broedersz,
Rep. Prog. Phys. \textbf{81}, 066601 (2018).

\bibitem{Edgar} \'E. Rold\'an,  et al., New J. Phys. {\bf 23}, 083013 (2021).

\bibitem{spiechowicz13} J. Spiechowicz, J {\L}uczka, and P. H{\"a}nggi, J.
Stat. Mech. \textbf{2013}, P02044 (2013).

\bibitem{Basu20} U. Basu, S. N. Majumdar, A. Rosso, S. Sabhapandit, and G.
Schehr, J. Phys. A \textbf{53}, 09LT01 (2020).

\bibitem{Marathe18} A. Saha, R. Marathe, and A. M. Jayannavar. J. Stat. Mech.
\textbf{2018}, 113203 (2018).

\bibitem{Marathe19} A. Saha and R. Marathe, J. Stat. Mech. \textbf{2019}, 094012
(2019).

\bibitem{Marathe22} R. Majumdar, A. Saha, and R. Marathe, J. Stat. Mech.
\textbf{2022}, 073206 (2022).

\bibitem{Ruben23} C. A. Guevara-Valadez, R. Marathe, and J. R. G{\'o}mez-Solano,
Physica A \textbf{609}, 128342 (2023).

\bibitem{Abhishek21} D. Chaudhuri and A. Dhar, J. Stat. Mech. \textbf{2021},
013207 (2021).

\bibitem{Lucente23} D. Lucente, A. Puglisi, M. Viale, and A. Vulpiani,
Phys. Rev. Lett. {\bf 131} (7), 078201 (2023).

\bibitem {Shee22} A. Shee and D. Chaudhuri, Phys. Rev. E \textbf{105}, 054148
(2022)

\bibitem{lenerev} K. N{\o}rregaard, R. Metzler, C. Ritter, K. Berg-S{\o}rensen,
and L. Oddershede, Chem. Rev. \textbf{117}, 4342 (2017).

\bibitem{lene1} J-.H. Jeon, N. Leijnse, L. Oddershede, and R. Metzler,
New J. Phys. \textbf{15}, 045011 (2013).

\bibitem{Blickle12} V. Blickle and C. Bechinger, Nat. Phys. \textbf{8}, 143
(2012).

\bibitem{Martinez16} I. A. Mart{\'i}nez, {\'E}. Rold{\'a}n, L. Dinis, D. Petrov,
J. M. R. Parrondo, and R. A. Rica, Nat. Phys. \textbf{12}, 67 (2016).

\bibitem{Wu20} X. L. Wu and A. Libchaber, Phys. Rev. Lett. \textbf{84}, 3017
(2000).

\bibitem{Krishnamurty21} N. Roy, N. Leroux, A. K. Sood, and R. Ganapathy,
Nat. Commun. \textbf{12}, 4927 (2021).

\bibitem{Albay21} J. A. C. Albay, Z.-Y. Zhou, C.-H. Chang, and Y. Jun, Sci.
Rep. \textbf{11}, 4394 (2021).

\bibitem{Cheng22} K. Cheng, P. Liu, M. Yang, and M. Hou, Soft Matter \textbf{18},
2541 (2022).

\bibitem{jaeoh} J. Shin, A. G. Cherstvy, W. K. Kim and R. Metzler,
New J. Phys. \textbf{17}, 113008 (2015).

\bibitem{activebath2} K. C. Leptos, J. S. Guasto, J. P. Gollub, A. I. Pesci,
and R. E. Goldstein,
Phys. Rev. Lett. \textbf{103}, 198103 (2009).

\bibitem{lowen} X. Zheng, B. Hagen, A. Kaiser, M. Wu, H. Cui, Z. Silber-Li
and H. L{\"o}wen, Phys. Rev. E \textbf{88}, 032304 (2013).

\bibitem{carsten} A. Cherstvy, O. Nagel, C. Beta, and R. Metzler,
Phys. Chem. Chem. Phys. \textbf{20} (2018).

\bibitem{runtumble} B. Partridge, S. Gonzalez Anton, R. Khorshed, G. Adams,
C. Pospori, C. Lo Celso, and C. F. Lee, PLoS ONE \textbf{17}, e0272587 (2022).

\bibitem{hapca} S. Hapca, J. Crawford, and I. Young,
J. Roy. Soc. Interface \textbf{6}, 111 (2008).

\bibitem{klapp} S. M. J. Khadem, N. H. Siboni, and S. H. L. Klapp,
Phys. Rev. E \textbf{104}, 064615 (2021).

\bibitem{elisabeth} E. Lemaitre, I. M. Sokolov, R. Metzler, and A. V. Chechkin,
New J. Phys. \textbf{25}, 013010 (2023).

\bibitem{Fox86} R. F. Fox, Phys. Rev. A \textbf{33}, 467-476 (1986).

\bibitem{Jung87} P. Jung and P. H{\"a}nggi, Phys. Rev. A \textbf{35},
4464-4466 (1987).

\bibitem{Cates09} J. Tailleur and M. E. Cates, EPL \textbf{86}, 60002 (2009)

\bibitem{Maggi15} C. Maggi, U. M. B. Marconi, N. Gnan, and R. Di Leonardo,
Sci Rep \textbf{5}, 10742 (2015).

\bibitem{Aswin21} A. Gopal, {\'E}. Rold{\'a}n, and S. Ruffo, J. Phys. A
\textbf{54}, 164001 (2021).

\bibitem{Fodor16} {\'E}. Fodor, C. Nardini, M. E. Cates, J. Tailleur, P. Visco,
and F. van Wijland, Phys. Rev. Lett. \textbf{117}, 038103 (2016).

\bibitem{Pak20} J. T. Park, G. Paneru, C. Kwon, S. Granick, and  H. K. Pak,
Soft Matter \textbf{16}, 8122-8127 (2020).

\bibitem{hanggi78} P. H{\"a}nggi, Z. Phys. B \textbf{30}, 85 (1978).

\bibitem{hanggi80} P. H{\"a}nggi, Z. Phys. B \textbf{36}, 271 (1980).

\bibitem{eliazar_klafter} I. Eliazar, J. Klafter, J. Stat. Phys. \textbf{111}, 739–768 (2003).

\bibitem{vandenbroeck} C. van den Broeck, J. Stat. Phys. \textbf{31}, 467-483 (1983).

\bibitem{spiechowicz14} J. Spiechowicz, P. H{\"a}nggi, and J. {\L}uczka, Phys.
Rev. E \textbf{90}, 032104 (2014).

\bibitem{spiechowicz15} J. Spiechowicz and J. {\L}uczka, Phys. Scr.
\textbf{2015}, 014015 (2015).

\bibitem{bialas23} K. Bia{\l}as, J. {\L}uczka, and J. Spiechowicz, Phys. Rev. E
\textbf{107}, 024107 (2023).

\bibitem{kanazawaPRL} K. Kanazawa, T.G. Sano, T. Sagawa, and H. Hayakawa, Phys. Rev. Lett. 114, 090601 (2015).

\bibitem{kanazawaJSP} K. Kanazawa, T.G. Sano, T. Sagawa, and H. Hayakawa, J. Stat. Phys. 160, 1294 (2015).

\bibitem{bialas20} K. Bia{\l}as, J. {\L}uczka, P. H{\"a}nggi, and J. Spiechowicz, Phys. Rev. E \textbf{102}, 042121 (2020).

\bibitem{baule} A. Baule and P. Sollich, Sci Rep. \textbf{13}, 3853 (2023).

\bibitem{prx} A. V. Chechkin, F. Seno, R. Metzler, and I. M. Sokolov, Phys. Rev. X \textbf{7}, 021002 (2017).

\bibitem{numerical07} C. Kim, E. Kyun Lee, P. H{\"a}nggi, and P. Talkner, Phys. Rev. E \textbf{76}, 011109 (2007).

\bibitem{numerical09} M. Grigoriu, Phys. Rev. E \textbf{80}, 026704 (2009).

\bibitem{statistics} M. B. Miller, Mathematics and statistics for financial risk management, 2nd Edition (Wiley, New York, 2013). 

\bibitem{hanggi2} P. H\"anggi,  Z.  Phys. B  {\bf 75} (2)  275  (1989).

\bibitem{baratoprl} A. Barato and U. Seifert, Phys. Rev. Lett. \textbf{115},
188103 (2015).

\bibitem{baratojpa} T. Wampler and A. Barato, J. Phys. A \textbf{55}, 014002
(2021).

\bibitem{Viktor20} V. Holubec, S. Steffenoni, G. Falasco, and K. Kroy, Phys.
Rev.  Res. \textbf{2}, 043262 (2020).

\bibitem{Viktor_Rahul_20} V. Holubec and R. Marathe, Phys. Rev. E \textbf{102},
060101 (2020).

\bibitem{Zakine2017} R. Zakine, A. Solon, T. Gingrich, and F. van Wijland, Entropy \textbf{19}, 193, (2017).


\bibitem{Onsanger1} L. Onsanger, Phys. Rev. 37, 405 (1930).

\bibitem{Onsanger2} L. Onsanger, Phys. Rev. 38, 2265 (1931).

\bibitem{crisantiJPA} A. Crisanti, F. Ritort, J. Phys. A: Math. Gen. 36 R181 (2003).

\bibitem{Julia} J. Bezanson, A. Edelman, S. Karpinski, and V. B. Shah,
SIAM Rev. \textbf{59}, 1 (2017).

\bibitem{Roldan2010} \'E. Rold\'an, and J. M. R. Parrondo, Phys. Rev. Lett.
\textbf{105}, 150607 (2010).

\bibitem{Steinberg} I. Z. Steinberg, Biophys J. \textbf{50}, 171-179 (1986).

\bibitem{Ohga} N. Ohga, S. Ito and A. Kolchinsky, Phys. Rev. E \textbf{131},
077101 (2023).

\bibitem{Shiraishi} N. Shiraishi, E-print arXiv:2304.12775.

\bibitem{Vu} T. V. Vu, V. T. Vo, and K. Saito, E-print arXiv:2305.18000.

\bibitem{liang2023thermodynamic} S. Liang and S. Pigolotti, E-print
arXiv:2308.14497.

\bibitem{Oberreiter} L. Oberreiter, U. Seifert, and A. C. Barato, Phys. Rev. E
\textbf{106}, 014106 (2022).

\bibitem{tobias}  T. Guggenberger, A. Chechkin, and R. Metzler,
J. Phys. A \textbf{54}, 29LT01 (2021).

\bibitem{tobias1} T. Guggenberger, A. V. Chechkin, and R. Metzler,
New J. Phys. \textbf{24}, 073006 (2022).

\bibitem{vojta} T. Vojta, S. Halladay, S. Skinner, S. Janu\v{s}onis, T.
Guggenberger, and R. Metzler, Phys. Rev. E \textbf{102}, 032108 (2020).

\bibitem{simulations} P. E. Kloeden , E. Platen, Numerical Solution of Stochastic Differential Equations (Springer Berlin, Heidelberg, 2013).



\end{thebibliography}

\appendix

\section*{Appendix}

\section{Moments of \texorpdfstring{$X_2$}{TEXT}}
\label{app:moments_X2}

We here prove Eqs.~\eqref{eqn:mean_X2}, \eqref{eqn:central_moments_X2_2_3}, and
\eqref{eqn:central_moments_X2_4} introduced in the main text. They can all be
straightforwardly proven using identities \eqref{eqn:moments_identity} and
\eqref{eqn:central_moments_general}. Since those two identities involve
derivatives of $I(q)$ evaluated at $k=0$, it will be convenient to find
first a general expression for $\frac{d^n}{dq^n}I(q)\rvert_{q=0}$.

\subsection{Derivatives of \texorpdfstring{$I(q)$}{TEXT}}

Derivatives of $I(q)$ can be easily obtained by means of the Taylor series.
First, from the definition of $I(q)$, Eq.~\eqref{eqn:I_definition}, it is
clear that $I(0)=0$. Equation \eqref{eqn:I_definition} can also be used to
find all derivatives of $I(q)$ denoted by $I^{(n)}(0)$. Let us start from
the Taylor expansion of $\hat{\rho}_a(q)$ around $k=0$,
\begin{equation}
\hat{\rho}_a(q)=\sum_{n=0}^{\infty}\dfrac{1}{n!}\hat{\rho}^{(n)}_a(0)q^n,
\end{equation}
where $\hat{\rho}^{(n)}_a$ denotes the $n$th derivative of $\hat{\rho}_a$.
The last expression substituted into \eqref{eqn:I_definition} gives the
first derivative of $I(q)$,
\begin{equation}
I'(q)=\dfrac{\hat{\rho}_a(q)-\hat{\rho}_a(0)}{q}=\sum_{n=1}^{\infty}\dfrac{
1}{n!}\hat{\rho}_a^{(n)}(0)q^{n-1}.
\end{equation}
Taking $m-1$ derivatives in $k$ we obtain
\begin{equation}
\dfrac{d^m}{dq^m}I(q)=\sum_{n=m}^{\infty}\dfrac{(n-1)(n-2)\ldots(n-m+1)}{n!}
\hat{\rho}_a^{(n)}(0)q^{n-m},
\end{equation}
which, evaluated at $q=0$, yields
\begin{equation}
I^{(m)}(0)=\dfrac{1}{m}\hat{\rho}_a^{(m)}(0)\mbox{ for }m\geq1.
\end{equation}
The moments of $\rho_a$ are related to the derivatives of $\hat{\rho}_a$
through the relation $\langle Y^m\rangle_a=\frac{1}{i^m}\hat{\rho}^{(m)}(0)$.
Hence
\begin{equation}
\label{eqn:I_derivatives}
I^{(m)}(0)=\dfrac{i^m}{m}\langle Y^m\rangle_a.
\end{equation}
We will make use of this result in the next subsection.

\subsection{Moments of \texorpdfstring{$X_2$}{TEXT}}
\label{momentX2}

We nowprove Eqs.~\eqref{eqn:mean_X2}, \eqref{eqn:central_moments_X2_2_3}, and
\eqref{eqn:central_moments_X2_4}. For the first we immediately see that
\begin{equation}
\langle X_2\rangle=\dfrac{1}{i}\dfrac{d}{dk}\hat{P}_2(q)\biggr\rvert_{q=0}=
\dfrac{1}{i}\omega\tau I'(0)=\omega\tau\langle Y\rangle_a,
\end{equation}
where we use expression \eqref{eqn:p1_p2_def} for $\hat{P}_2(q)$ and
Eq.~\eqref{eqn:I_derivatives}. To prove Eqs.~\eqref{eqn:central_moments_X2_2_3}
and \eqref{eqn:central_moments_X2_4} for simplicity we define $f(q)\equiv I(q)-
qI'(0)$; according to Eq.~\eqref{eqn:central_moments_general} we need to take
the derivatives of $\exp{(\omega\tau f(q))}$. The second derivative evaluated
at $q=0$ reads
\begin{equation}
\dfrac{d^2}{dq^2}\exp{(\omega\tau f(q))}\rvert_{q=0}=\left[\left(\omega^2\tau^2
f'(q)^2+\omega\tau f''(q)\right)\exp{(\omega\tau f(q))}\right]\rvert_{q=0}=
\omega\tau I^{(2)}(0),
\end{equation}
since $f(0)=f'(0)=0$ and $f^{(n)}(0)=I^{(n)}(0)$. Using this expression along
with Eqs.~\eqref{eqn:central_moments_general} and \eqref{eqn:I_derivatives} we
obtain for the second moment
\begin{equation}
\langle(X_2-\langle X_2\rangle)^2\rangle=\dfrac{1}{2}\omega\tau\langle Y^2
\rangle_a,
\end{equation}
which proves Eq.~\eqref{eqn:central_moments_X2_2_3}, where $n=2$. Analogously,
now consider the third derivative of $\exp{(\omega\tau f(q))}$,
\begin{equation}
\dfrac{d^3}{dq^3}\exp{(\omega\tau f(q))}\rvert_{q=0}=\left[\left(\omega^3\tau^3
f'(q)^3+3\omega^2\tau^2f'(q)f''(q)+\omega\tau f'''(q)\right)\exp{(\omega\tau
f(q))}\right]\rvert_{q=0}=\omega\tau I^{(3)}(0).
\end{equation}
Thus the third central moment of $X_2$ reads
\begin{equation}
\langle(X_2-\langle X_2\rangle)^3\rangle=\dfrac{1}{3}\omega\tau\langle Y^3
\rangle_a,
\end{equation}
proving Eq.~\eqref{eqn:central_moments_X2_2_3}, where $n=3$. To prove
Eq.~\eqref{eqn:central_moments_X2_4} we consider the fourth derivative
\begin{multline}
\dfrac{d^4}{dq^4}\exp{(\omega\tau f(q))}\rvert_{q=0}=\biggr[\big(\omega^4
\tau^4f'(q)^4+6\omega^3\tau^3f'(q)^2f''(q)+3\omega^2\tau^2f''(q)^2+\\
4\omega\tau^2f^{(3)}(q)f'(q)+\omega\tau f^{(4)}(q)\big)\exp{(\omega\tau f(q))}
\biggr]\biggr\rvert_{q=0}=3\omega^2\tau^2\left(I^{(2)}(0)\right)^2+\omega\tau
I^{(4)}(0).
\end{multline}
Hence, we have
\begin{equation}
\langle(X_2-\langle X_2\rangle)^4\rangle=\dfrac{3}{4}\omega^2\tau^2\langle Y^2
\rangle_a^2+\dfrac{1}{4}\omega\tau\langle Y^4\rangle_a,
\end{equation}
Which completes our proof.

\section{Time-dependent moments}
In most of our work,  we have just considered the moments of the stationary distribution. Since the time-dependent distribution is not available, we do not have access to the time evolution of the moments of the position. Nevertheless, starting from the Fokker-Planck equation, we can construct a set of ordinary differential equations for the finite-time moments $\langle X_t^n\rangle$ for $n \in \mathbb{N}$. By multiplying Eq.~\eqref{eqn:FokkerPlanck} by $x^n$ and integrating over $x$ we obtain
\begin{equation}
    \tau \dfrac{\partial}{\partial t} \langle X_t^n \rangle = -n \langle X_t^n \rangle + \dfrac{k_BT}{\kappa} n(n-1) \langle X_t^{n-2} \rangle + \omega \tau \left( \langle (X_t+Y)^n \rangle - \langle X_t^n\rangle \right), \label{eq:costantinoNov}
\end{equation}
with the convention that $\langle X_t^{-1} \rangle=0$, and 
\begin{equation}
    \langle ( X_t+Y)^n \rangle = \int_{-\infty}^{\infty} \int_{-\infty}^{\infty} (x+y)^n P(x,t) \rho_a(y) dxdy.
\end{equation}
Equation~\eqref{eq:costantinoNov} can be rewritten using the binomial formula
\begin{equation}
    \tau \dfrac{\partial}{\partial t} \langle X_t^n \rangle = -n \langle X_t^n \rangle + \dfrac{k_BT}{\kappa} n(n-1) \langle X_t^{n-2} \rangle + \omega \tau \sum_{m=0}^{n-1} \binom{n}{m} \langle X_t^m \rangle \langle Y^{n-m} \rangle_a.
    \label{eq:costantinoNov2}
\end{equation}
Therefore, we find a hierarchy of relations for the generic $n$-th moment in terms of the first to the $(n-1)$-th moment. A general analytical solution to Eq.~\eqref{eq:costantinoNov2} is not straightforward, nevertheless equations governing the first two moments are relatively simple
\begin{align}
    \tau \dfrac{\partial}{\partial t} \langle X_t \rangle &= -\langle X_t \rangle + \omega \tau \langle Y \rangle_a,\\
    \tau \dfrac{\partial}{\partial t} \langle X_t^2 \rangle &= -2 \langle X_t^2 \rangle + 2\dfrac{k_BT}{\kappa} + \omega\tau \langle Y^2 \rangle_a + 2\omega \tau \langle X_t \rangle \langle Y \rangle_a,
\end{align}
thus the first moment reads
\begin{equation}
    \langle X_t \rangle = \langle X_0 \rangle \exp(-t/\tau) + \left( 1- \exp(-t/\tau) \right) \omega \tau \langle Y \rangle_a
\end{equation}
while the second moment
\begin{eqnarray}
    \langle X_t^2 \rangle&=& \langle X_0^2 \rangle e^{-2t/\tau} + \left( \dfrac{k_BT}{\kappa} + \dfrac{\omega\tau }{2} \langle Y^2 \rangle_a \right) \left[1- \exp(-t/\tau) \right]  \\
   &+& 2\omega \tau \langle Y \rangle_a \langle X_0 \rangle \left[ \exp(-t/\tau) - \exp(-2t/\tau) \right] + \omega^2 \tau^2 \langle Y \rangle_a^2 \left[\exp(-2t/\tau) -2\exp(-t/\tau) +1 \right].
\end{eqnarray}
It is then possible to obtain the variance which reads
\begin{equation}
    \mathrm{Var}[X_t] = \left( \dfrac{k_BT}{\kappa} + \dfrac{\omega\tau }{2} \langle Y^2 \rangle_a \right) \left[ 1- \exp(-t/\tau) \right].
\end{equation}
Despite the presence of the Poissonian shot noise, the system relaxes to its stationary state with a characteristic time $\tau$ that does not depend on the rate of the kicks.

\section{Derivation of susceptibility}
We will here expose the analytical derivation of the function $\chi_X(t)$ \eqref{eqn:chi_def}. 
The average of $\langle X_t \rangle_{\delta f}$ can be obtained with a minimal modification of equation \eqref{eq:costantinoNov}
\begin{equation}
    \tau \dfrac{\partial}{\partial t} \langle X_t \rangle_{\delta f} = -\langle X_t\rangle_{\delta f} + \dfrac{\delta f}{\kappa} + \omega \tau \langle Y \rangle_a.
\end{equation}
Clearly, since the perturbation is independent from the noise $\eta_t$, the moments of $Y$ remain unaltered. The susceptibility will then read
\begin{equation}
    \chi_X(t) = \dfrac{1}{\kappa} \left( 1- \exp(-t/\tau) \right),
\end{equation}
which at stationarity becomes 
\begin{equation}
    \chi_X = \dfrac{1}{\kappa}.
\end{equation}

\section{Numerical simulations}
\label{sec:code}
We conclude the paper with a short discussion on how to simulate the system. Trajectories evolving with the Langevin equation can be generated via the Euler-Maruyama scheme described in any textbook on numerical implementation of stochastic differential equations (we refer to \cite{simulations}). While the analytical expression \eqref{eqn:stat_P} can be simulated using the GNU Scientific Library (GSL) available in many languages (we used the Julia language
\cite{Julia}).
To calculate Eq.~\eqref{eqn:stat_P} we proceeded in three steps:
(i) we compute the function $I(q)$ in Eq.~\eqref{eqn:I_definition} for a discrete set of values of $q$; (ii) we interpolate the resulting values with a cubic spline interpolation available in the package \texttt{Interpolations.jl} in order to find a continuous version of $I(q)$; (iii) we plug this function into formula \eqref{eqn:stat_P} and integrate for a range of values of $x$. All numerical integrations were performed using the Gauss-Kronrod algorithm available in the package \texttt{QuadGK.jl}. We  simulate the Langevin system \ref{eqn:Langevin} using the first-order integrator method. The Gaussian noise $\xi_t$ is derived from the Wiener process, and the active kicks are generated at Poisson-distributed times with rate $\omega$.

\end{document}